\documentclass[final,3p,times,twocolumn]{elsarticle}

\usepackage{amsmath,amssymb}

\usepackage{xcolor}

\usepackage{ulem}
\usepackage[hidelinks]{hyperref}
\usepackage{bm}

\journal{Powder Technology}

\begin{document}
\begin{frontmatter}

\title{Influence of embedding depth on the pullout of a plate from a confined granular medium}

\author[inst1]{Dominik Krengel\corref{mycorrespondingauthor}}
\ead{dominik.krengel@kaiyodai.ac.jp}
\cortext[mycorrespondingauthor]{Corresponding author}
\affiliation[inst1]{organization={Department of Marine Resources and Energy, Tokyo University of Marine Science and Technology},
            addressline={4-5-7 Konan}, 
            city={Minato},
            postcode={108-8477}, 
            country={Japan}}

\author[inst2]{Jian Chen}
\affiliation[inst2]{organization={Center for Mathematical Science and Advanced Technology, Japan Agency for Marine-Earth Science and Technology},
            addressline={3173-25, Showa-machi, Kanazawa-ku}, 
            city={Yokohama},
            postcode={236-0001}, 
            country={Japan}}

\author[inst1]{Shun Nomura}
\author[inst1]{Hidenori Takahashi}

\begin{abstract} 
 Understanding the behaviour of moving objects within confined granular media is important in many technical applications. In this work we study the vertical pullout of a plate-like intruder buried at different depths within a dense granular aggregate from the onset of motion to flow around the intruder in the steady state by means of a two-dimensional polygonal Discrete Element Method. 
We find that the peak and steady-state drag forces depend differently on their embedding depth for different friction coefficients. Depending on the strength of friction the plate may interact with the domain walls, which enhances the drag force until failure of the aggregate, but not in the steady-state as dilation around intruder directs the force chains emanating from the plate upwards and away from the walls. If the inter-particle friction is weak, it is possible for the overhead pressure to overcome frictional resistance beyond a certain embedding depth. This prevents the formation of dilated failure zones above the intruder and changes the failure mode of the aggregate, leading to a slower decay of the steady-state drag forces with pullout distance.
\end{abstract}



\begin{keyword}
Intruder pullout\sep Stress-Strain behaviour\sep Micromechanics\sep Shape evolution\sep Discrete-Element Simulation
\end{keyword}

\end{frontmatter}

\section{Introduction}\label{sec:introduction}
The mechanics of an object moving through or interacting with a granular medium has been studied for decades across multiple disciplines, such as geomechanics\,\cite{Liu2012,Sivakumar2013,Liang2021, Kurniadi2025, Nomura2026,Krengel2026}, industrial manufactoring\,\cite{Radl2012, Boonkanokwong2016, Lehuen2020, Nadeem2021}, or even bio-inspired locomotion in terramechanics\,\cite{Li2012,Li2013}. 

In many instances, the intruder is initially at rest, held in place through contacts with surrounding particles; the surrounding medium is a dense as particle shape, material properties permit and loading history permit. Once the object begins to move, whether through applied forces or with prescribed velocity, it experiences an increasing load, the drag force, from the resistance of the surrounding granular matrix to the deformation induced by the intruder motion. At a material-dependent threshold force, the material in the direction of motion fails, i.e. it dilates and the drag force on the intruder rapidly decreases. After some time, the material has maximally dilated and the drag force remains constant.  

This steady-state is typically for granular fluids\,\cite{Albert1999,Albert2001,Henann2014,Seguin2016,Takada2019,He2025}. Depending on the intruder velocity, the rearrangement of particles around the intruder changes, leading to different,  velocity-dependent drag force regimes\,\cite{Wassgren2003, Potiguar2013}. At low intruder velocities, in the quasi-static regime, the response to the intruder motion is velocity independent\,\cite{Hilton2013} and instead depends on the overburden pressure of by the particles above the object, i.e. its depth. This leads to a number of differences whether the object is pulled horizontally, or vertically against the direction of gravity. In the prior case, the overhead pressure remains constant, whereas in the latter case the overhead pressure decreases as the intruder is moving upwards. 

The additional resistance of the aggregate to the intruder movement due to embedding depth can lead to different, depth-dependent failure modes\,\cite{Jalali2021}, which geotechnical literature distinguishes as deep and shallow failures\,\cite{Das2013}, although categorising the ultimate drag force with respect to the embedding depth, normalized by the area of the object\,\cite{Fujikata2026} implies that the distinction is not as strict as previously stated. Experiments have led to various semi-analytical formulations to predict the peak drag force at failure (e.g.\,\cite{Trautmann1985,Murray1987, Tagaya1988}), based on different assumed shapes of the failure surface around the intruder, like outward facing, curved failure surfaces\,\cite{White2008} or inward facing conical failure surfaces for deep embedded intruders\,\cite{Liang2021}, which are different stages in the evolution of the failure during the motion of the intruder\,\cite{Krengel2026a}. Further research also found dependencies of the failure surface at different stages of the pullout on a combination of the embedding depth and local packing density\,\cite{Liu2012,Evans2019,Kurniadi2025}, the intruder shape\,\cite{Khatri2011,Askari2016}, size of the granular particles\,\cite{Costantino2008, Athani2017,Lehuen2020}, and to increase with larger friction coefficients $\mu$\cite{Meyerhof1968,Rowe1982,Merifield2006,Shahriar2020}, as the contacts between particles on the edge of the uplift area can support more weight from above with less support from below.

The immediate force response and evolution of the contact networks prior to failure also depend on the initial depth of the intruder\,\cite{Shah2020} and the distance between the intruder and the boundaries of the granular domain. Strong force networks reaching the containing walls of the aggregate enhance jamming of particles between the intruder and the wall and so increase the drag forces acting on the intruder\,\cite{Carvalho2022}. Making and breaking of these strong force chains within the contact network then leads to likewise strong fluctuations in the drag forces\,\cite{Geng2005,Zhang2017,Basak2023}, which increase the deeper the intruder is buried in the aggregate\,\cite{Hossain2020a}. In the steady-state instead, the force chains tend to concentrate in dense, conical clusters ahead of the intruder\,\cite{Lehuen2020, Kozlowski2021, Krengel2026a}, rather than connecting to the walls of the granular domain, which further leads to a strong localisation  of the granular flows around the intruder\,\cite{Seguin2016}

Despite significant advances, a unified understanding of the micro- and macromechanical responses to an intruder moving through a granular medium is still severely lacking. The majority of works on intruder pullout has focused on the initial failure itself, instead of the steady state regime, or even the whole process. Complications also arise in particular from the presence of domain walls that may affect the intruder and its depth below the free surface. In our preceding work\,\cite{Krengel2026a} we studied the pullout of a plate-like intruder from a granular medium placed at one specific depth below the surface. We showed the emergence of an effective shape in the steady state, independent of the frictional properties of the aggregate, that determines its steady-state characteristics but not the ultimate drag force acting on the plate. 
In this study, we extend our previous work, to further clarify the influence of initial embedding depth on the micro- and macroscopic behaviour of the plate and aggregate, and how friction and depth together shape the influence of the walls on the intruder kinematics throughout the pullout process. Our results are presented in terms of the initial depth of the plate rather than its current depth, as the plate motion disturbs the aggregate; two plates at the same depth beneath the surface may have entirely different ambient conditions depending how long the plate has already moved, and so can no longer be directly compared.

\section{DEM Simulation}
\label{sec:simulation}
We conduct several Discrete Element (DEM) simulations to investigate the mechanical response of the aggregate to the pullout of the intruder. The simulation algorithm is described in detail in\,\ref{app:DEM}. Although realistically the pullout process in a three-dimensional medium, we instead conduct two dimensional simulations so that our systems can be considered a two-dimensional slice through the real setup, similar to previous quasi-2D experiments using discs (e.g.\,\cite{Kozlowski2021, Zhang2017, Kozlowski2021, Carvalho2022}). The results for the 2D system qualitatively agree with results obtained from experiments as shown previously in\,\cite{Krengel2026}. Using a two-dimensional geometry allows a considerable speedup of the simulation for equivalent system sizes, in particular for the non-spherical particles used in our simulations. 

\begin{figure}[t!]
 \centering
 \includegraphics[width=\columnwidth]{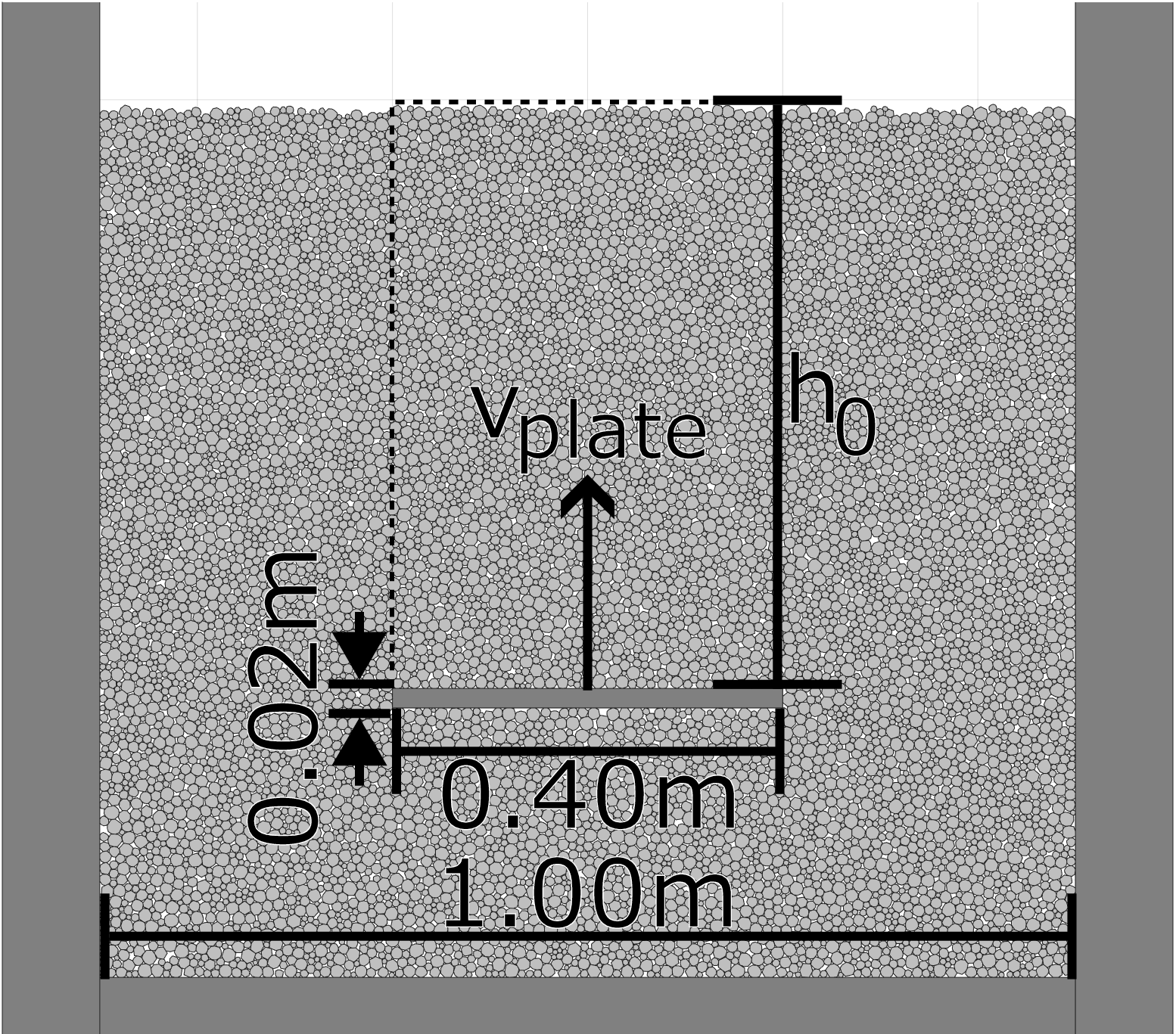}
 \caption{Initial configuration of the pullout test. The lateral walls and the bottom plate remain fixed while the plate is slowly pulled out with constant velocity $v_{\mathrm{plate}}$ from different initial embedding depths $h_{0}$.}
 \label{fig:pullout_setup}
\end{figure}

Particles are modelled as irregular, convex polygons with 6 to 13 corners with a uniform circum-radius distribution. The mean radius of the particles is $\bar{r}=0.005$ m and varies by up to 50\%. The intruder is modelled as a rectangular polygon with sidelength $0.4$\,m by $0.05$\,m or approximately 40 by 2 particle diameters. Figure\,\ref{fig:pullout_setup} shows the simulation setup. The particle size affects the absolute values of the forces acting on the intruder, however the qualitative trends remain the same\,\cite[Fig. A.17]{Krengel2026}. Here, we chose a particle size small enough to allow a sufficient number of particles on the intruder for force evaluation, but permit fast execution times of the simulation. 
 
We generate the initial deposit by first dropping 3000 frictionless particles into an empty box under gravity. The configuration is smoothed and the plate placed is central on top of the particle layer. Then more particles are deposited on top of the plate until the box is filled. The aggregate is then trimmed to the desired embedding depth and left to equilibrate, during which the deposit slightly compacts or expands, depending on the specific overhead load. The resulting deviation from the targeted filling height is less than 1\% for each assembly generated. We consider five different initial embedding depths for the intruder, $h_{0} = [0.20,0.40,0.60,0.78,1.20,1.63]$\,m, covering a wide range of depth to system width ratios. For each embedding depth, except $h_0=1.63$\,m, we generate three different samples and present the obtained results as sample averages. For $h_0=1.63$\,m we chose to simulate only one sample to confirm the behaviour shown by $h_0=1.2$\,m.

After deposition friction is enabled and the system left to equilibrate.  The plate is then slowly lifted upwards with constant velocity $v_{\mathrm{plate}}=0.01$\,m/s for 20 seconds. During pullout the plate is unaffected by the material response and only moves upwards without horizontal movement or rotation. The plate velocity is slow enough that the resulting deformation is quasi-static; doubling the plate velocity will lead to the same stress response. For each embedding depth and sample we simulate the pullout with three different friction coefficients $\mu=[0.1, 0.5, 1.0]$ for exactly the same initial state each. Further simulation parameters are given in tab.\,\ref{tab:materialparameters}. As the simulations are performed in two dimensions, the dimensions of the material density $\varrho$ and the Young modulus $Y$ are given in units of [kg/m$^2$] and [N/m] to allow definition of the force in units of [N]. Assuming a unit thickness in the out-of-plane direction also recovers the corresponding three dimensional units for $\varrho$ and $Y$.

\begin{table}[h!]
 \caption{Numerical parameters for the simulations.}
 \begin{tabular}{lll}
Parameter & & Value\\
\hline
 Num. particles  & $N$   & $\sim 12000$\\
 Coeff. friction & $\mu$ &  $[0.1,0.5,1.0]$\\
 Young modulus   & $Y$   & $10^{9}$ [N/m]\\
 Damping parameter & $\gamma$ & $0.5$\\
 Material density & $\varrho$ & $5000$ [kg/m$^2$]\\
 Mean particle radius & $\bar{r}$ & $0.005$ [m]\\
 Particle diameter range & & $[0.5\ 1.5] \bar{r}$\\
 Plate size      & $l_{\mathrm{plate}}\times d_{\mathrm{plate}}$ & $0.4\times0.02$ [m]\\
                 &                                       & $\widehat{=}80\times4$ [$\bar{r}$]\\
 Plate depth     & $h_0$ & $[0.20,0.40,$\\
                 &       & $0.60,0.78,$ \\
                 &       & $1.20,1.63]$ [m]\\
 Pullout velocity & $v_{\mathrm{plate}}$ & $0.01$ [m/s]\\
                  &                      & $\widehat{=} 2$ [$\bar{r}$/s]\\
 Timestep        & $\mathrm{d}t$ & $2.5e-6$ [s]\\
 Simulation time & $t_{\mathrm{max}}$ & 20 [s]\\
\hline
 \end{tabular}
 \label{tab:materialparameters}
\end{table}

\section{Results}\label{sec:results}
\subsection{Dragforce}
\begin{figure}[b!]
 \centering
 \includegraphics[width=\columnwidth]{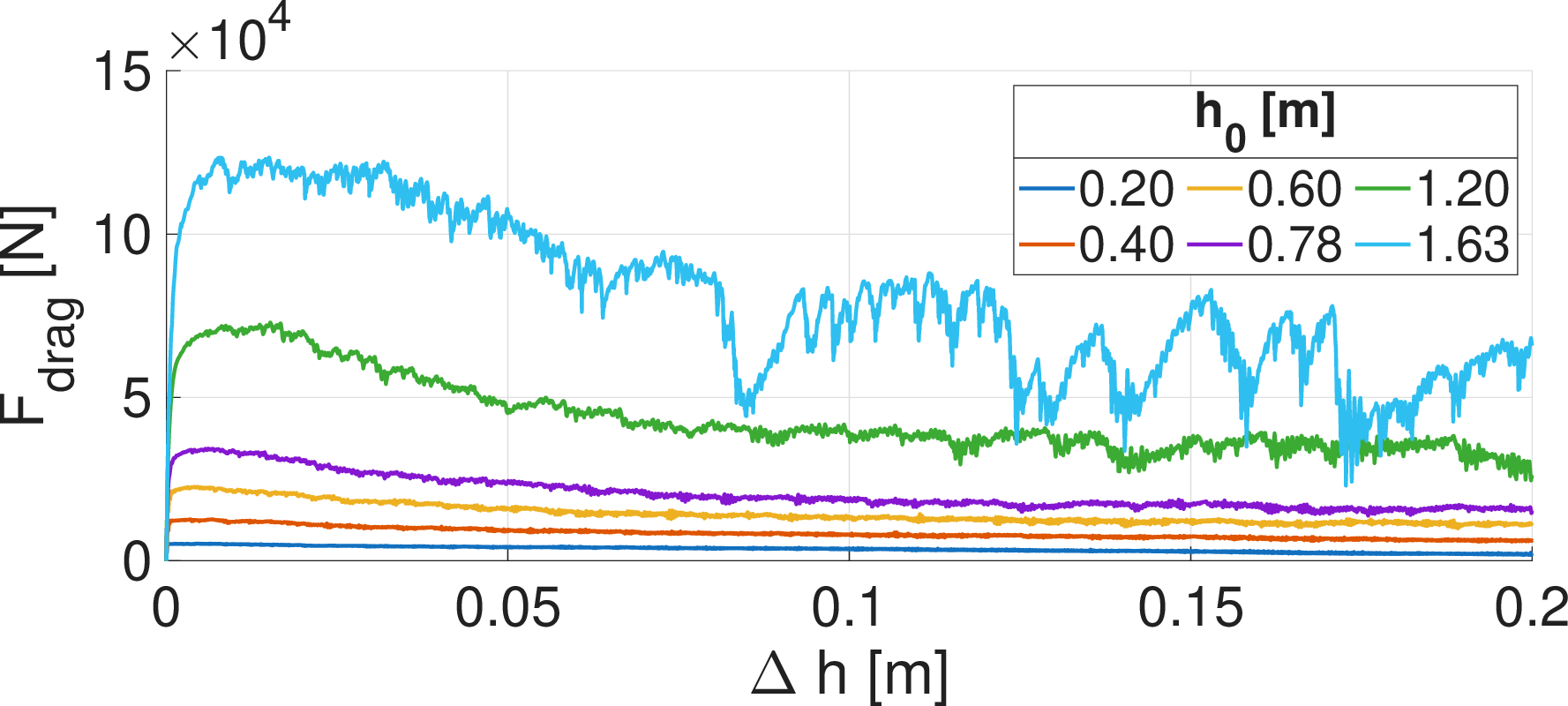}
 \caption{Evolution of the drag force $F$ with increasing displacement $\Delta h$ depending on the initial depth of the plate, for $\mu=0.5$.}
 \label{fig:dragforce_dh_mu}
\end{figure}
Figure\,\ref{fig:dragforce_dh_mu} shows the evolution of drag force $F_{\mathrm{drag}}$ with increasing displacement $\Delta h=v_{\mathrm{plate}}\cdot t$ for different initial embedding depth $h_{0}$ and $\mu=0.5$. When we move the plate upwards the drag force first increases as the surrounding granular matrix hardens in response to the loading. Then the force peaks and begins to decrease, at first non-linear, then linear as the material has reached a steady-state equivalent with linearly decreasing confining pressure. 
Increasing the depth of the plate leads to a stronger confinement forces and and increases the overall force response on the plate. Increasing the strength of friction forces between the particles and between the particles and the plate increases the peak drag force, but leads to the same steady state drag forces. For $\mu=0.1$, the clear drag-force peak almost disappears, indicative of the failure of weakly frictional materials\,\cite{Krengel2025}: the peak broadens significantly before a small non-linear decrease marks the onset of the steady state regime.
Regardless of embedding depth, the drag force shows small, high frequency fluctuations superimposed on strong, only partially regularly occurring drops in force (see $F_{\mathrm{drag}}$ for $h_0 = 1.63$\,m in Fig.\,\ref{fig:dragforce_dh_mu} for an especially pronounced example). The strength of both types of fluctuations, relative to the force average, increase in magnitude as the initial embedding depth increases. In contrast to experiments\,\cite{Hossain2020,Fujikata2026}, we observe the strongest fluctuations in our two-dimensional simulations in the steady-state rather than in the preceding non-linear decay phase, which matches the force fluctuations reported for two-dimensional simulations with discs reported in\,\cite[Fig. 1c)]{Athani2019}. We conjecture that this is an influence of dimensionality, as in three dimensions, particles have more possibilities to stick, slip, and jam, and the failure surface around the intruder will be more complicated compared to the two dimensional case investigated here. Our current data is insufficient to confirm our hypothesis, however we are planning to address this issue in a follow-up study.

The displacement $\Delta h$ used here is relative, not the absolute depth of the plate below the surface. This means, the load on the plate and its associated micromechanics cannot be stated in terms of a simple depth-relation, but need to instead consider the initial depth of the intruder. In other words: At the same depth below the surface of the aggregate, the drag force and the micro-mechanics for two plates in the same assembly can be entirely different, depending how far they have already moved upwards.

\begin{figure}[t!]
 \centering
 \includegraphics[width=\columnwidth]{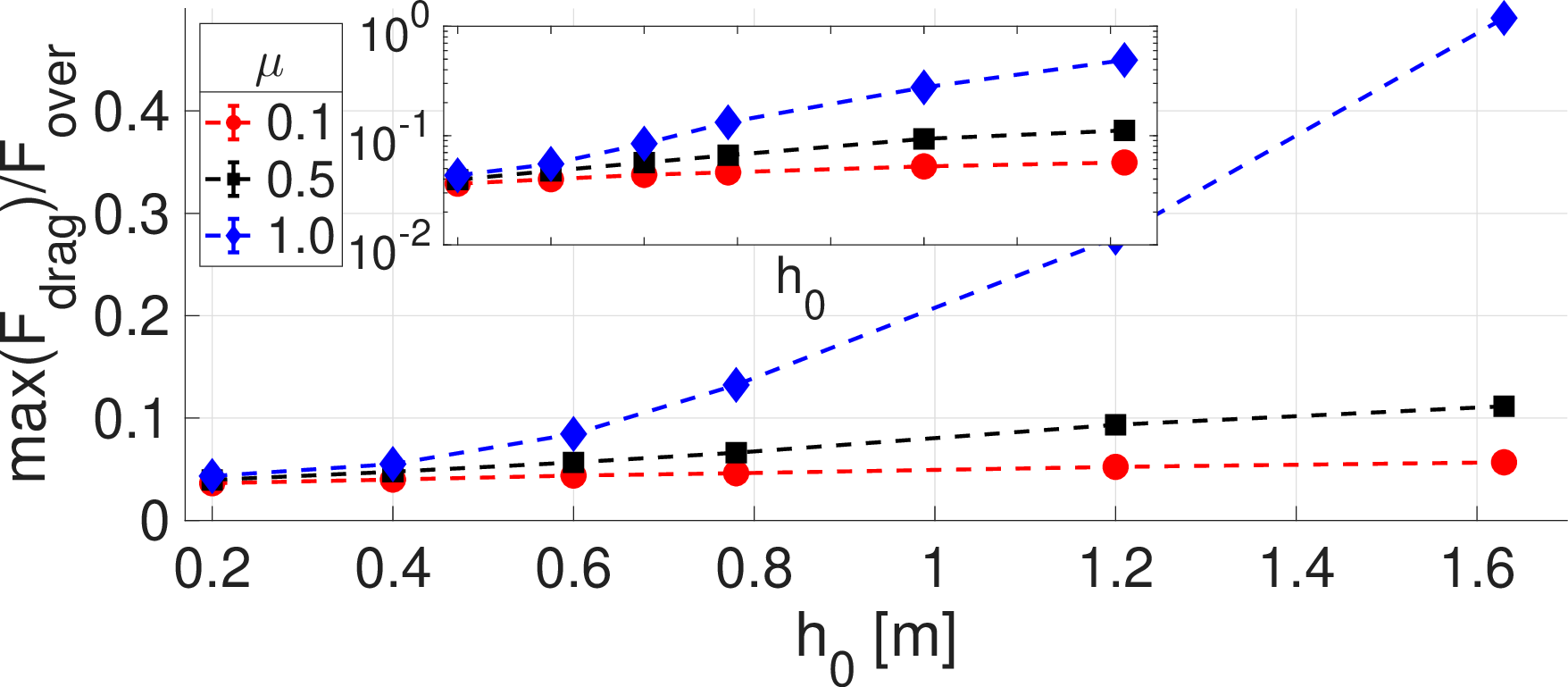}
 \caption{Dependence of the maximal drag force on embedding depth $h_0$ and friction coefficient $\mu$, scaled with the initial overburden force $F_{\mathrm{over}}$. The inset shows the same data in a semi-logarithmic plot. The data points represent averages over three samples each. The errorbars are too small to be seen at this scale.}
 \label{fig:maxdragforce_dh_mu}
\end{figure}

In Fig.\,\ref{fig:maxdragforce_dh_mu} we show the maximum drag force $\mathrm{max}(F_{\mathrm{drag}})$, scaled with the weight of particles directly above the plate, $F_{\mathrm{over}}$ (marked with dashed lines in Fig.\,\ref{fig:pullout_setup}). For all values of $\mu$, $\mathrm{max}(F_{\mathrm{drag}})$ increases non-linear with $h_0$. Larger $\mu$ increases the influence of $h_0$: The granular matrix is both more resistant to reorganization and a higher overburden pressure from deeper embedding can also mobilize more friction. The functional dependence of the maximum drag force on $h_0$ is sub-exponential (see also the inset in Fig.\,\ref{fig:maxdragforce_dh_mu}), with the saturation of the drag-force setting in earlier for weaker friction coefficients. For shallower embedding and larger friction coefficients the functional form of our observed maximum drag force also matches the maximal drag scaling reported by\,\cite[Fig. 4]{Shah2020} for different system and particle sizes.

\begin{figure}[b!]
 \centering
 \includegraphics[width=\columnwidth]{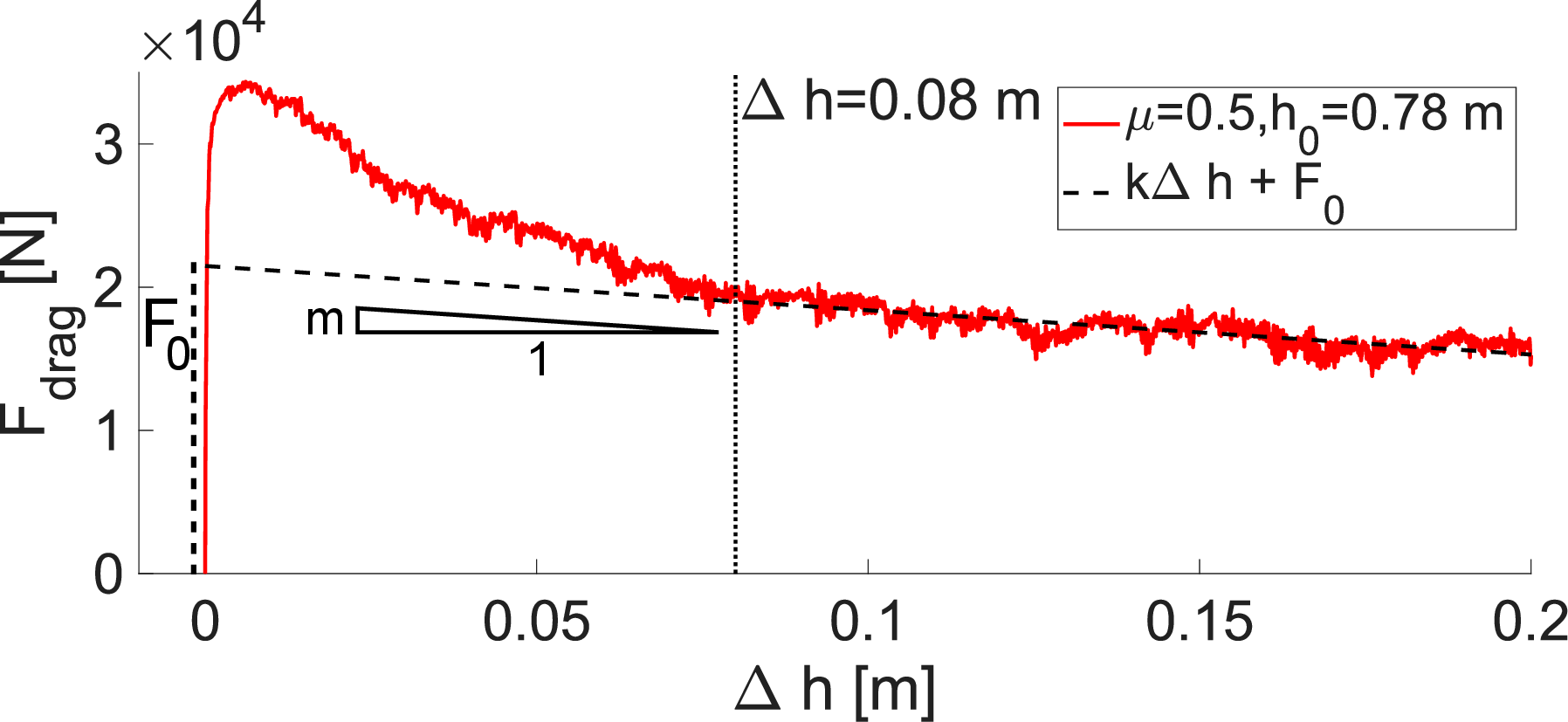}
 \caption{Example of the linear fit to the steady-state drag force showing the slope $m$ and the intercept $F_0$.}
 \label{fig:dragforce_dh_fit}
\end{figure}

\begin{figure}[b!]
 \centering
 \includegraphics[width=\columnwidth]{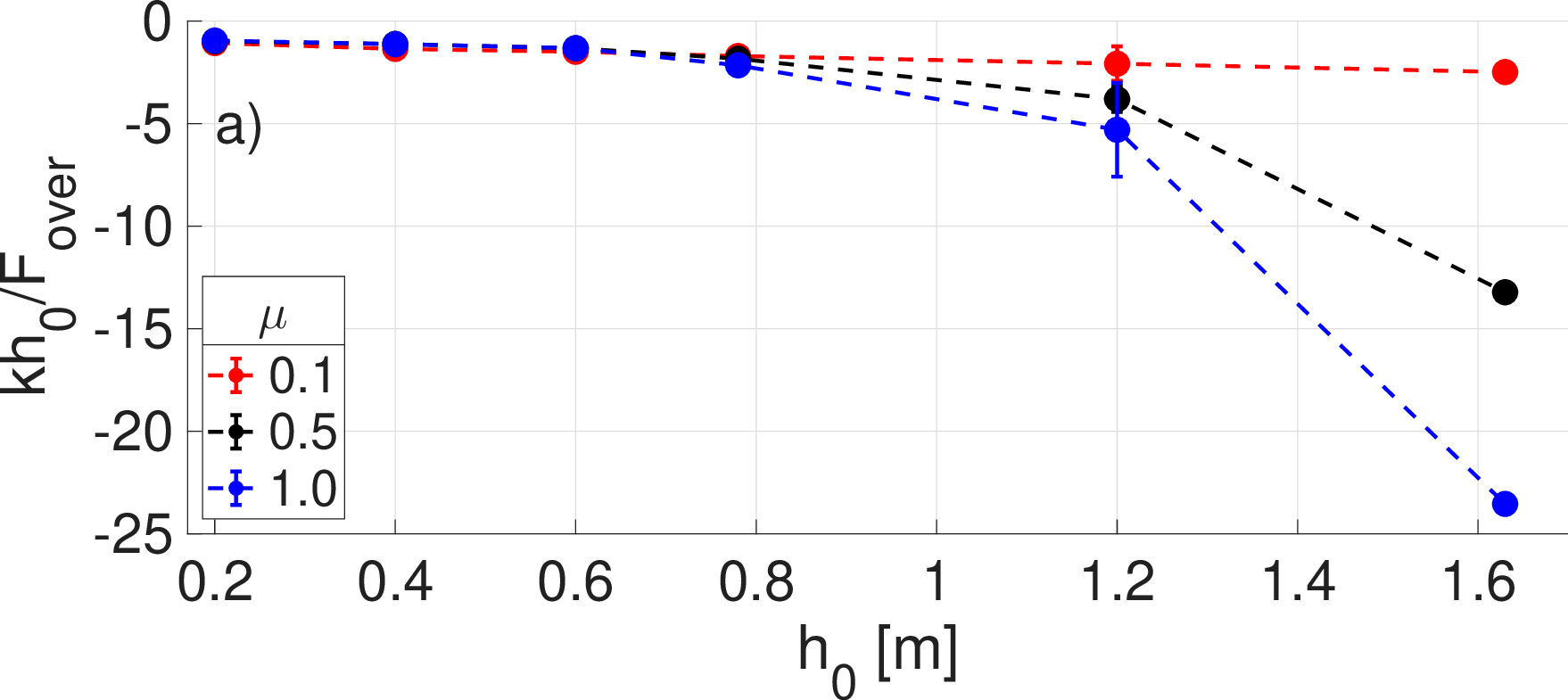}

 \includegraphics[width=\columnwidth]{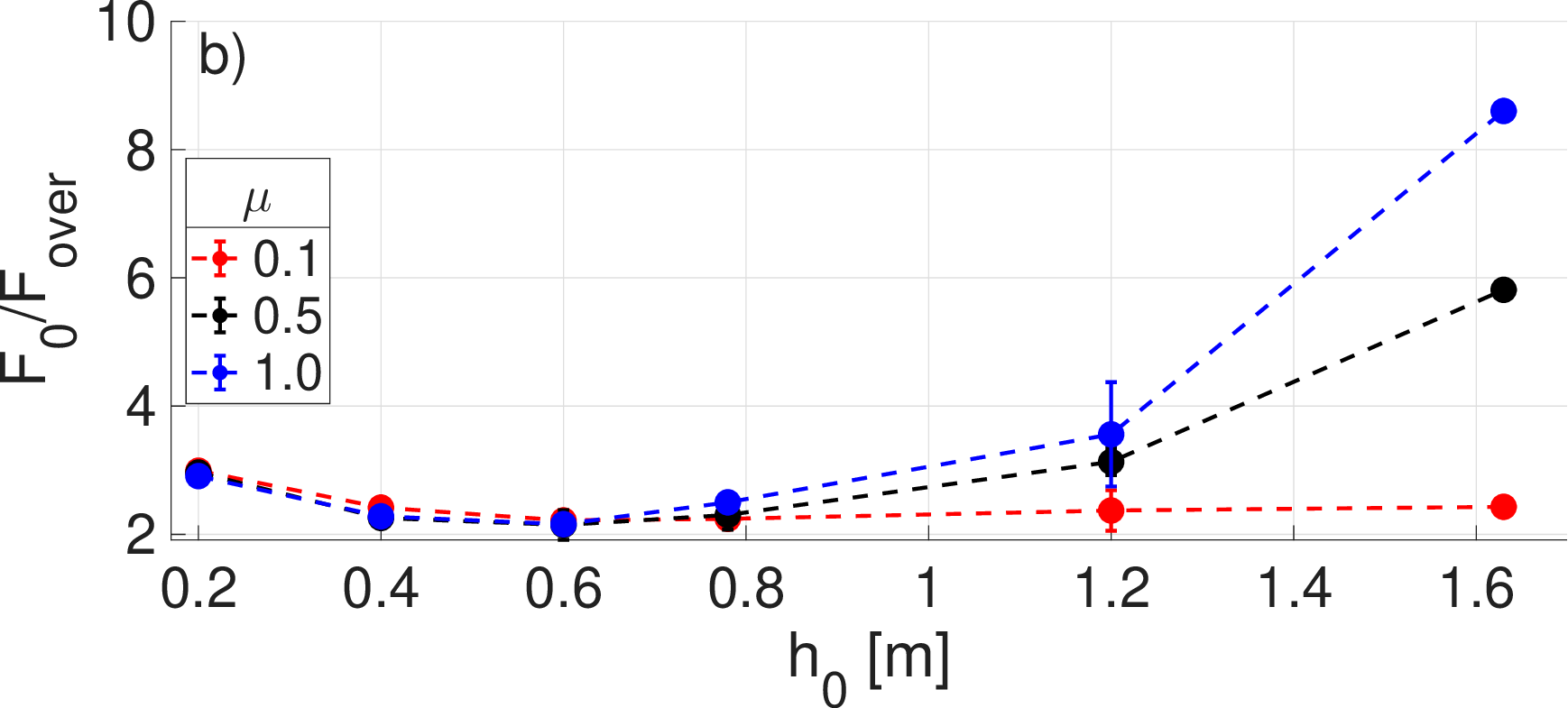}
 \caption{Functional dependence of the slope (a) and the intercept (b) fit parameter on the initial embedding depth $h_0$. Both parameters are scaled to dimensionless values. All data points, except for $h_0=1.63$\,m are sample averages. The errorbars are typically to small to see at this scale.}
 \label{fig:dh_F_fit}
\end{figure}

We also perform a linear fit,
\begin{equation}
 F_{\mathrm{drag}}(\Delta h\geq0.08) = k\Delta h + F_0,
\end{equation}
to the steady-state forces, starting from $\Delta h = 0.08$\,m (Fig.\,\ref{fig:dragforce_dh_fit}). While lower values of $\Delta h$ are also viable choices to start the fit from for some configurations, we chose $\Delta h=0.8$ as starting point as from there on all configurations had reliably entered the steady state regime. As the fluctuations in the drag force increase with embedding depth and friction coefficient, the quality of the fit decreases correspondingly, from an $R^2 = 0.98$ at $h_0=0.2$\,m and $\mu=0.1$ to $R^2 = 0.51$ at $h_0=0.2$\,m and $\mu=1.0$. Physically, we can understand the slope of the fit $k$, i.e. the rate with which $F_{\mathrm{drag}}$ decreases, as the residual stiffness of the granular aggregate above the plate after failure. The intercept $F_0$, on the other hand, gives a rough estimate of the ultimate drag force of an already established granular flow where no dilation needs to occur. Figure\,\ref{fig:dh_F_fit} shows both fit parameters normalized with an initial overburden force on the plate,
\begin{equation}
 F_{\mathrm{over}} = M_{\mathrm{p}}\cdot g,
\end{equation}
where $g=9.81$ is the gravitational acceleration and $M_{\mathrm{p}}$ the total particle mass in a rectangular area from the plate to the free surface, see Fig.\,\ref{fig:pullout_setup}. The slope $k$ of the fit is further scaled with the embedding depth to nondimensionalize it.

 For shallow embedding depths, $h_0\leq 0.6$\,m, $k$ is barely dependent on $\mu$ and scales linear with $h_0$, see Fig.\,\ref{fig:dh_F_fit}, a). For $h_0=0.78$\,m differences start to emerge in $k$ between the different materials, where $k(\mu_1)<l(\mu_2)$ for $\mu_1>\mu_2$. For $h_0\geq1.2$\,m the influence of $\mu$ become more pronounced. We further find that the dependence of $k$ on $h_0$ changes: For $\mu=0.1$ the linear dependence continues, but for $\mu\geq0.5$ $k$ clearly scales non-linear with $h_0$. While we cannot exclude that $k(\mu=0.1)$ will also show non-linear behaviour outside our investigated range of $h_0$, it still indicates that some differences occur in the failure mechanism for plates embedded deeply in weakly frictional aggregates compared to embedding in strongly frictional materials. 
In contrast to the monotonous decrease of $k$, the linear intercept $F_0$ is decreasing independently of $\mu$ for $h_0\leq 0.6$\,m, where it reaches a minimum. For $h_0\>0.6$\,m $F_0$ increases again, and friction becomes relevant, as the increase is faster if $\mu$ is larger, see Fig.\,\ref{fig:dh_F_fit}, b). 
For both $k$ and $F_0$, the variation between samples is negligible small at low embedding depth, but increases for larger $h_0$. That both parameters for the steady-state fit are almost independent of $\mu$ up to $h_0=0.6$\,m stands in strong contrast to the ultimate force, where the divergence happened at much shallower embedding. It also shows, that the force transmission following failure of the aggregate changes with the embedding depth of the intruder.

For the steady-state parameters, the divergence between the different friction branches occurs around $h_0 =0.78$\,m, while the peak force showed an earlier divergence between $0.40<h_0<0.60$\,m, which we understand as the divergence depth not being a strict singular value, but rather a transition regime in which different parameters react with different sensitivity to friction and embedding depth.

While we perform our simulations here with a constant ratio of plate size to system width, we also expect that changing the system width will affect the functional dependencies of $m$ and $F_0$: An increase in system width should make the influence of $h_0$ on $m$ and $F_0$ more evident in the current ``shallow'' region ($h_0\leq 0.8$\,m).

\begin{figure*}[t!]
 \includegraphics[width = \textwidth]{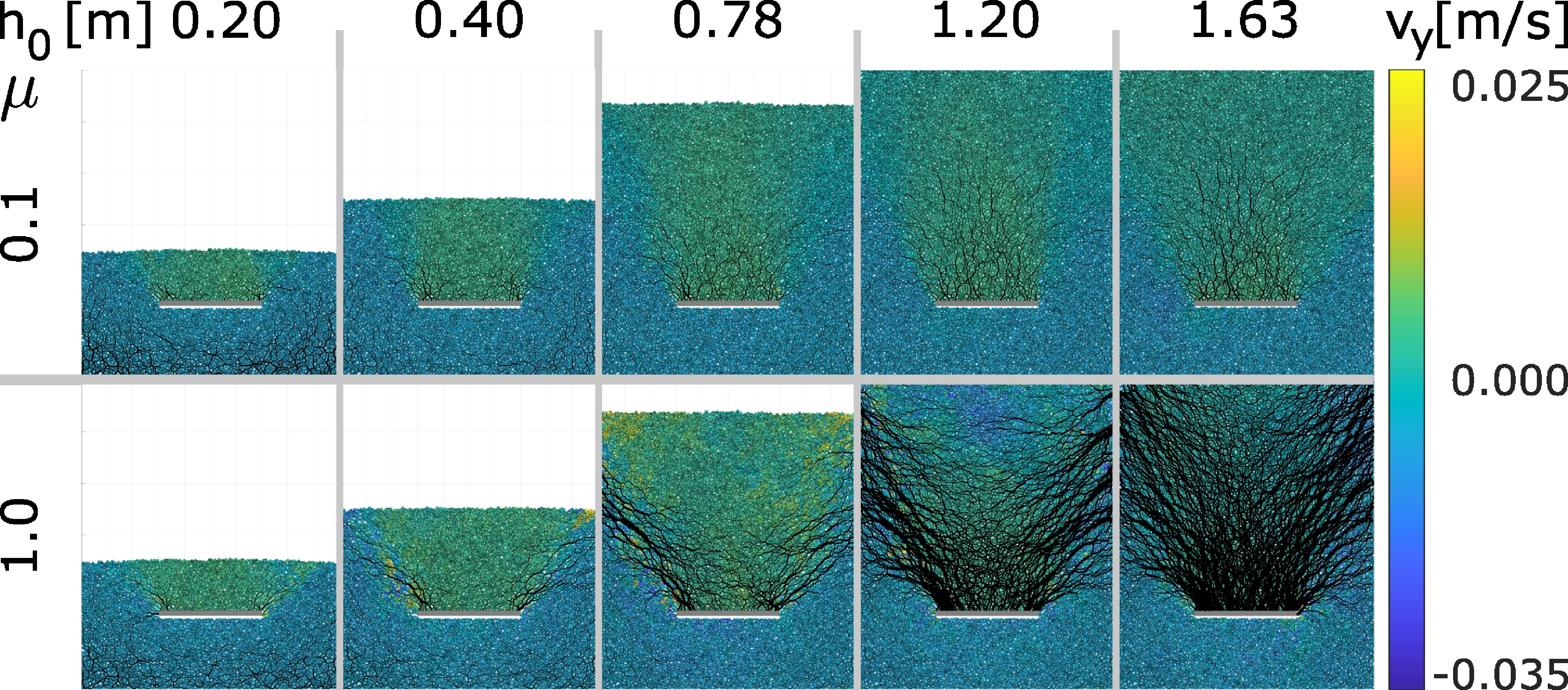}
 \caption{Force chain network $\propto F_{N}$ and relative vertical velocity $v_y$ at maximal drag force for different $h_0$ and $\mu$. Green particles move upwards at the same speed as the plate, yellow particles move faster, and blue particles move downwards relative to the plate. The line thickness of the force chains is normalized relative to total weight of particles directly above the plate. To maintain readability, the line strength for $\mu=0.5$ has been additionally scaled up with a factor of 0.5.}
 \label{fig:forcechain_velo_peak}
\end{figure*}

\subsection{Macroscale response}
\subsubsection{Forces and velocities}\label{sec:forcechains}
When the drag force peaks, the plate pushes a trapezodial wedge upwards (Fig.\,\ref{fig:forcechain_velo_peak}), with the failure surface, i.e. the boundary of the uplifted wedge,  pointing roughly 60$^{\circ}$. At shallow depths, the force chains attach to the plate primarily on the plate edges, i.e. close to where the surrounding granular material will fail. They are both stronger and longer, the deeper the plate is buried in the aggregate. For depths $h_0\geq 0.78$\,m  the force chains also begin to emanate from the entire plate surface, not only the edges.
Increasing the friction coefficient generally leads to the same behaviour, however a number of details differ: The failure surfaces now point stronger outwards, with an angle of about 45$^{\circ}$ for $\mu=1.0$, as contacts are generally more stable, and so more particles can be lifted by the plate. The overall strength of the force network also increases, and for $h_0\geq0.78$\,m strong force chains, significantly beyond the median force in the aggregate, begin to connect the plate with the domain walls.

\begin{figure}[b!]
 \centering
 \includegraphics[width = \columnwidth]{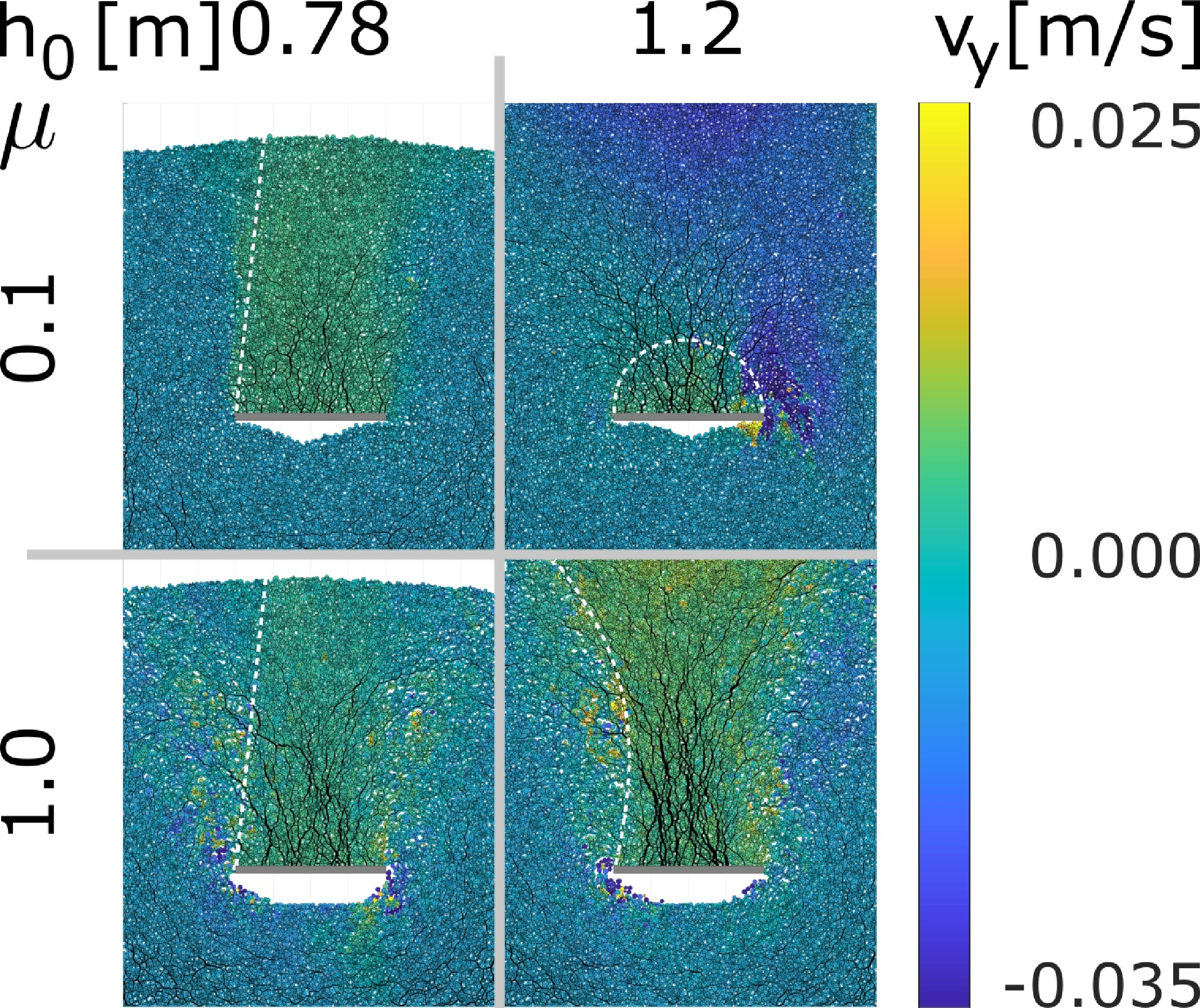}
 \caption{Force chain network $\propto F_{N}$ and relative vertical velocity $v_y$ at the onset of the steady state ($\Delta h=0.08$\,m) for different $h_0$ and $\mu$. Green particles move upwards at the same speed as the plate, yellow particles move faster, and blue particles move downwards relative to the plate. The line thickness of the force chains is normalized relative to total weight of particles directly above the plate. To maintain readability, the line strength for $\mu=0.5$ has been additionally scaled up with a factor of 0.5. The white lines mark the approximate shape of the failure surface along the uplift wedge. In the case of ($\Delta h=1.2$\,m, $\mu = 0.1$), no clear failure surface is observable; the white line only marks the boundary of the uplift wedge.}
 \label{fig:forcechain_velo_steady}
\end{figure}

At the onset of the steady state (Fig.\,\ref{fig:forcechain_velo_steady}) the force network and the velocity profile have changed significantly. The failure surfaces bounding the uplift wedge have expanded into dilated failure zones generally shaped like a downwards pointing triangle (see the guiding lines in Fig.\,\ref{fig:forcechain_velo_steady} for $\mu=1.0$ and $h_0 = 0.78$\,m). In these failure zones strong, intermittent downwards flow appears, which transports displaced particles from the surface to below the plate. As friction increases, the size of these failure zones also increases outwards towards the container walls.
The inner boundaries of the failure zone generally start to demark a stable triangular uplift area, centered around the middle of the plate\,\cite{Krengel2026}, which remains constant during further pullout. In the case of near surface embedding ($h_0=0.20$\,m), the inner failure surface remains pointed upwards, the triangular structure does not form, as there is not enough material and overhead pressure above the plate. 
For deep embedding, $h_0\geq1$\,m, the geometry of the uplift wedge changes significantly depending on $\mu$: For weak friction ($\mu=0.1$) the uplifted area recedes towards the plate, forming a roughly semispherical domain with strong fluctuations in its boundary, which matches the deep failure described in Geotechnical literature (e.g\,\cite{Das2013}). The force chains have slightly increased in strength compared to more shallow embedding and reach outside the uplift wedge. We observe no clear failure surface in this case.
At large friction ($\mu\geq0.5$), the uplift wedge still reaches the free surface. The bounding failure surfaces are curved, with a narrow waist around halfway its height. These curved failure surfaces are transient, with increasing plate displacement they will also turn to bound a triangular uplift wedge. For all embedding depths at $\mu=1.0$ the force chains are mostly contained within the uplift wedge and generally point in the direction of plate displacement. Increasing the embedding depth $h_0$ of the intruder increases their lengths, but barely affects their strength. Particle flow in the dilated zone and around the intruder is stronger compared to all cases for $\mu=0.1$. We additionally note that while the backfilling below the plate is primarily controlled by the friction coefficient, which determines the angle of repose, for $\mu=0.1$ increasing the embedding depth $h_0$ seems to additionally decrease the void space below the plate.

\subsubsection{Porosity}
\begin{figure}[t!]
 \centering
 \includegraphics[width=\columnwidth]{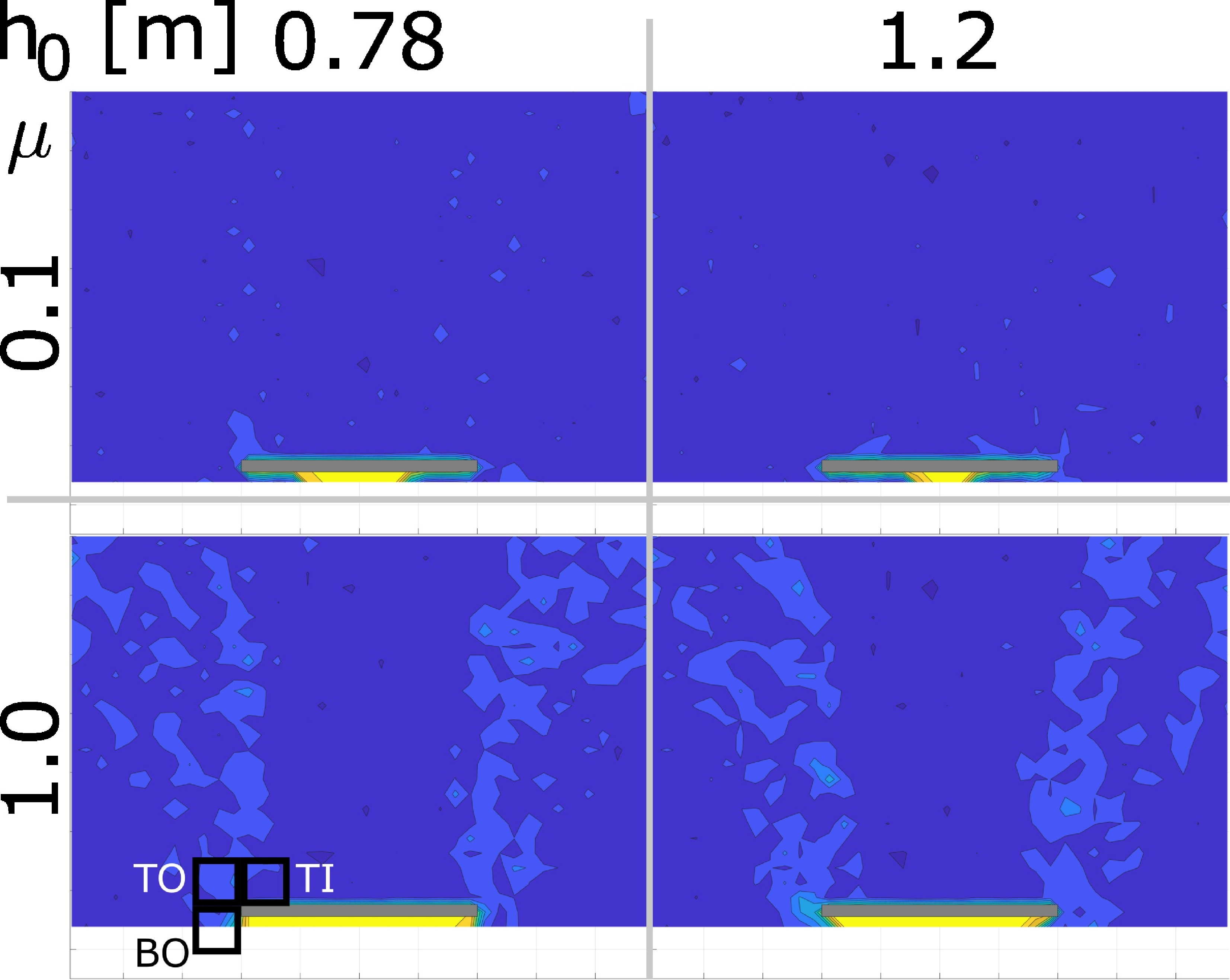}
 \caption{Porosity at the onset of the steady state ($\Delta h=0.08$\,m) for $h_{0}=0.78$\,m and $1.2$\,m at low and high friction. Lighter colouring indicates lower porosity, yellow indicates empty spaces below the plate. The boxes in the bottom left subfigure mark the location where we track the local fabric evolution. Box TO is located above the plate, outside the edge. Box TI is located above the plate, inside the edge. Box BO is located below the upper plate edge, outside the edge.}
 \label{fig:porosity}
\end{figure}
As the plate moves upwards, the granular matrix fails, and, as our packing is initially dense, the granulate should dilate in response to the plate movement. We compute the porosity field,
\begin{equation}
 n=1-\frac{\sum{A_{\mathrm{particles}}}}{A_{\mathrm{box}}},
\end{equation}
by subdividing the domain into rectangular boxes $A_{\mathrm{box}}=(2d_{\mathrm{plate}})^2$ and computing the total particle area $A_{\mathrm{particles}}$ within each box. 
Figure\,\ref{fig:porosity} shows that for $\mu=0.1$ a chain of low porosity zones extends upwards from the plate edges if $h_{0}\leq 0.78$\,m, following the boundaries of the co-moving wedge, i.e. the failure surface, shown in Fig.\,\ref{fig:forcechain_velo_steady}. Below this depth, $h_{0}> 0.78$\,m, no notable dilation occurs, the porosity above the plate effectively remains low and homogeneous after failure. While Fig.\,\ref{fig:forcechain_velo_steady} shows the formation of a semispherical uplift wedge, it is no longer coinciding with dilation. 
In contrast, strongly frictional systems ($\mu\geq0.5$) show the formation of an extensive dilation zone at all depths, confirming the previous observations, reaching from the edges of the plate to the free surface of the aggregate and to the walls of the container. Although $\mathrm{max}(F_{\mathrm{drag}})$ and $F_{0}$ directly increase with $\mu$, the absence of dilation in the low friction case explains why the steady-state drag force decreases at a lower rate $k$ (shown in Fig.\,\ref{fig:dh_F_fit}, b) compared to the high friction case: The aggregate is denser and therefore provides more residual resistance to the intruder motion.

That no dilation occurs for $\mu=0.1$ and $h_0\geq 1.20$\,m also suggests, that there is a critical depth, where the initial overhead pressure is enough to overcome the frictional resistance between the grains. Any dilation that occurs due to the plate movement is then immediately countered by recompaction of the aggregate due to the overhead pressure and the plate motion, which leads to very localized and compact failure around the plate. In our case, this critical depth lies between $h_0=0.78$\,m and $h_0=1.2$\,m. For larger $\mu$, this dilation may occur at deeper embedding, although only up to a limit. At some point, the friction forces should be large enough to completely halt translational movement, so that particles can only roll to escape the applying forces. The transition from a sliding-dominated motion to a rolling-dominated motion may further affect the evolution of failure around the intruder\,\cite{DeGiuli2016}. In this case, a deep-failure may not occur at all; however our current data is insufficient to support our speculation. The absence of dilation for ($h_0\geq1.20$\,m, $\mu=0.1$) also shows why the drag force not only scales with the current depth of the intruder, as shown earlier in Figs.\,\ref{fig:maxdragforce_dh_mu} and\,\ref{fig:dh_F_fit}, but also with its initial embedding depth: Unlike as for water, the granulate compacts, fails and dilates in response to the intruder motion, so its density is not only position- but also time-dependent. For an object at the same depth, the history of the system can thus lead to an entirely different dilation state and thus to entirely different forces. This may not be an issue for granular flows where a static intruder is placed within a particle flow, but it certainly affects the drag force when pulling an intruder from rest from within a granular medium.

\begin{figure}[t!]
 \centering
 \includegraphics[width=\columnwidth]{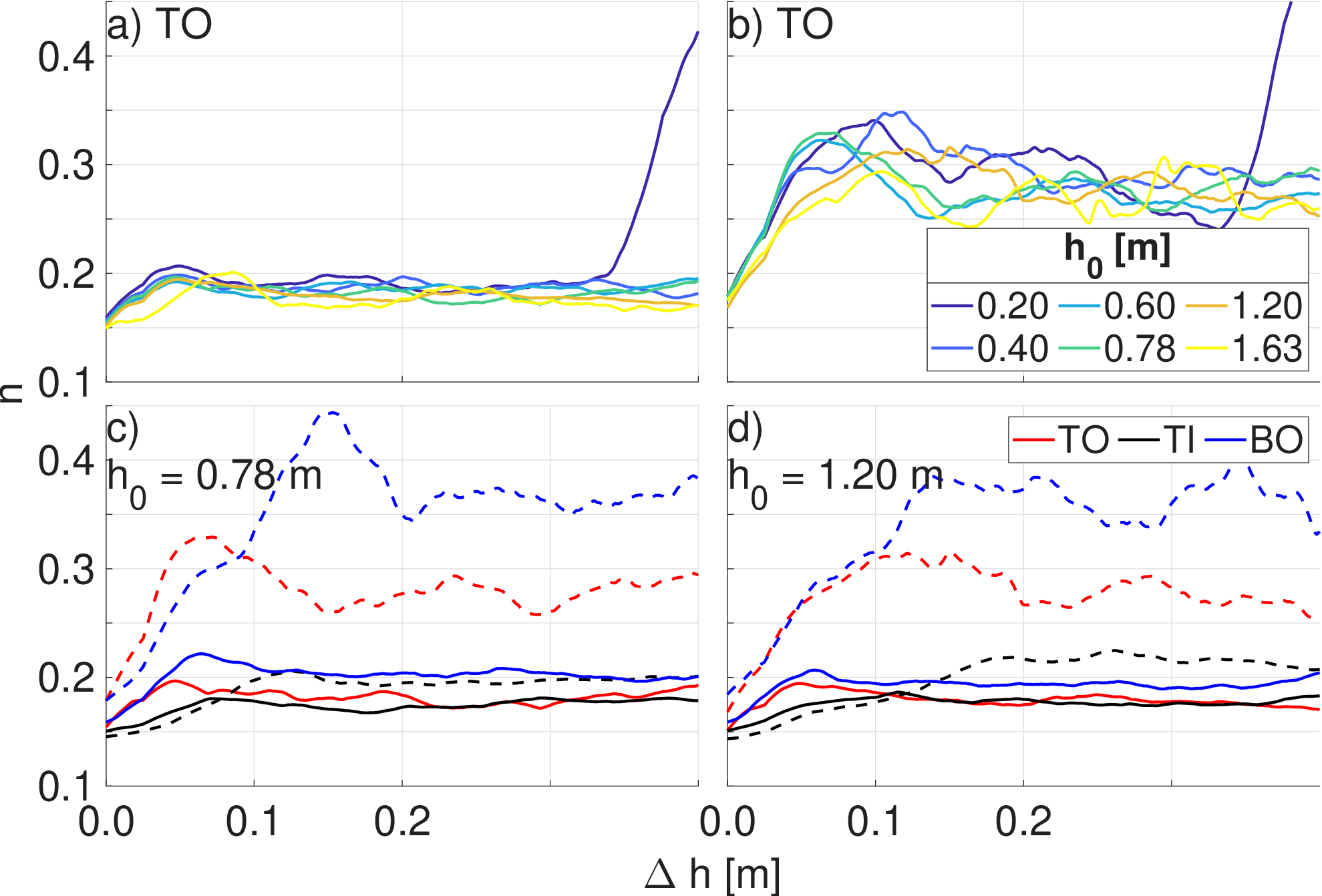}
 \caption{Evolution of the porosity around the plate: Immediately outside on top of the plate (TO) for a) $\mu=0.1$ and b) $\mu=1.0$ at different embedding depth $h_{0}$. c) For $h_{0}=0.78$\,m for TO, inside on top of the plate (TI) and outside below the plate (BO). Solid lines show $\mu=0.1$ and dashed lines $\mu=1.0$. d) The same as for c), with $h_{0}=1.20$\,m. See also Fig.,\ref{fig:porosity} for the location of the sampling areas. All data are presented as smoothed averages over 30 data points.}
 \label{fig:porosity_plate}
\end{figure}
As Fig.\,\ref{fig:porosity_plate} shows, for $\mu\geq0.05$ and $h_0\geq1.2$\,m the porosity through the granular domain remains largely unchanged, while for the other cases large dilation zones occur. In all cases, however, the aggregate dilates locally around the plate to permit particle transport from aboce to below the plate. To understand how the porosity around the plate evolves with the displacement for different combinations of $h_{0}$ and $\mu$ we select three different sampling areas around the plate: Areas immediately outside the edge, both above (TO) and below (BO) will react to the induced dilation and particle flow around the plate, while the sampling area inside the edge (TI) should react to the formation of the uplift wedge, see also Fig.\,\ref{fig:porosity} for a graphical representation of the location of the sampling areas. Each sampling location is square with a sidelength of 0.04\,m, i.e. twice the plate thickness. The sampling locations move upwards together with the plate. The resulting data is first averaged over the three different samples, then over the matching boxes left and right of the plate. 

Figure\,\ref{fig:porosity_plate} a) and b) show the porosity at TO for $\mu=0.1$ and $\mu=1.0$. For $h_{0}=0.20$\,m the porosity spikes as the plate nears the free surface since the sampling area now increasingly reaches outside the aggregate. Initially the porosity increases as the plate uplifts the particles in the higher layers, then the aggregate above the plate fails and particles flow again in the now opened space. The porosity after failure fluctuates around a higher level than it had initially, likely due to the aggregate being initialized without friction, but the pullout being conducted with a finite value for $\mu$. The situation remains the same if the value of $\mu$ increases, however now the peak porosity and steady state average are higher by around 50\% compared to $\mu=0.1$, and the magnitude of the fluctuations around the steady value increases. For $\mu=0.1$ the curves appear to collapse onto the same form. For $\mu=1.0$ the curves still follow the same overall form but vary more strongly. Fig.\,\ref{fig:porosity_plate}, a) may possibly indicate a decrease of the steady state porosity $n$ with embedding depth $h_0$ for $\mu=0.1$, our data is not clear enough to ascertain this observation. For $\mu=1.0$ the data is too noisy to indicate any possible dependence on $h_0$.

Generally, the qualitative and quantitative evolution of the local porosity around the plate is independent of the initial embedding depth $h_0$ and friction $\mu$ (Fig.\,\ref{fig:porosity_plate} c) and d). Outside the plate, both above (TO) and below (BO), $n$ increases at almost the same rate, but reaches a lower maximum above the plate than below: The strong force chains emanating from the plate (see also Fig.\,\ref{fig:forcechain_velo_peak}) lead to jamming of contacts, the formation of particle arches and so prevent the downflow of particles besides the plate before failure. This momentarily increases the effective plate width and leads to larger porosity below the plate. After failure of the granular matrix, the movement of the plate particle eventually breaks these arches and particles begin to flow down, decreasing the porosity below the plate. As it takes time for the plate movement to interrupt the jamming, the peak porosity below the plate occurs after the peak porosity above the plate. Increasing the friction coefficient leads to higher porosity and later occurrence of peak porosity as the arches are more stable. Inside the plate edges (TI), the increase in porosity is slower than outside the plate as the dilation follows the spread of the failure surface from initially outward pointing to the final triangular wedge. In the steady state, for $\mu=0.1$ the porosity above the plate converges against the same value, independent of $h_0$. Below the plate the porosity remains about 18\% larger. For large $\mu=1.0$, the porosity above the plate is significantly larger outside the plate than it is inside, as the aggregate inside the plate edges remains stable, but large friction prevents refilling of the dilated aggregate outside the plate. While the evolution of $n$ is largely independent of $h_0$, we observe one clear difference between deep and shallow failure for $\mu=1.0$: For $h_0\leq0.78$\,m $n$ shows a pronounced maximum in every location around the plate, i.e. the aggregate dilates and then compacts again. In contrast, for $h_0>0.78$\,m no such clear maximum exists. Instead the aggregate dilates to the same level as in the steady state of the shallow failure, and then remains at this value without any re-compaction.

\subsection{Evolution of the uplift wedge}
\begin{figure}[b]
 \centering
 \includegraphics[width=\columnwidth]{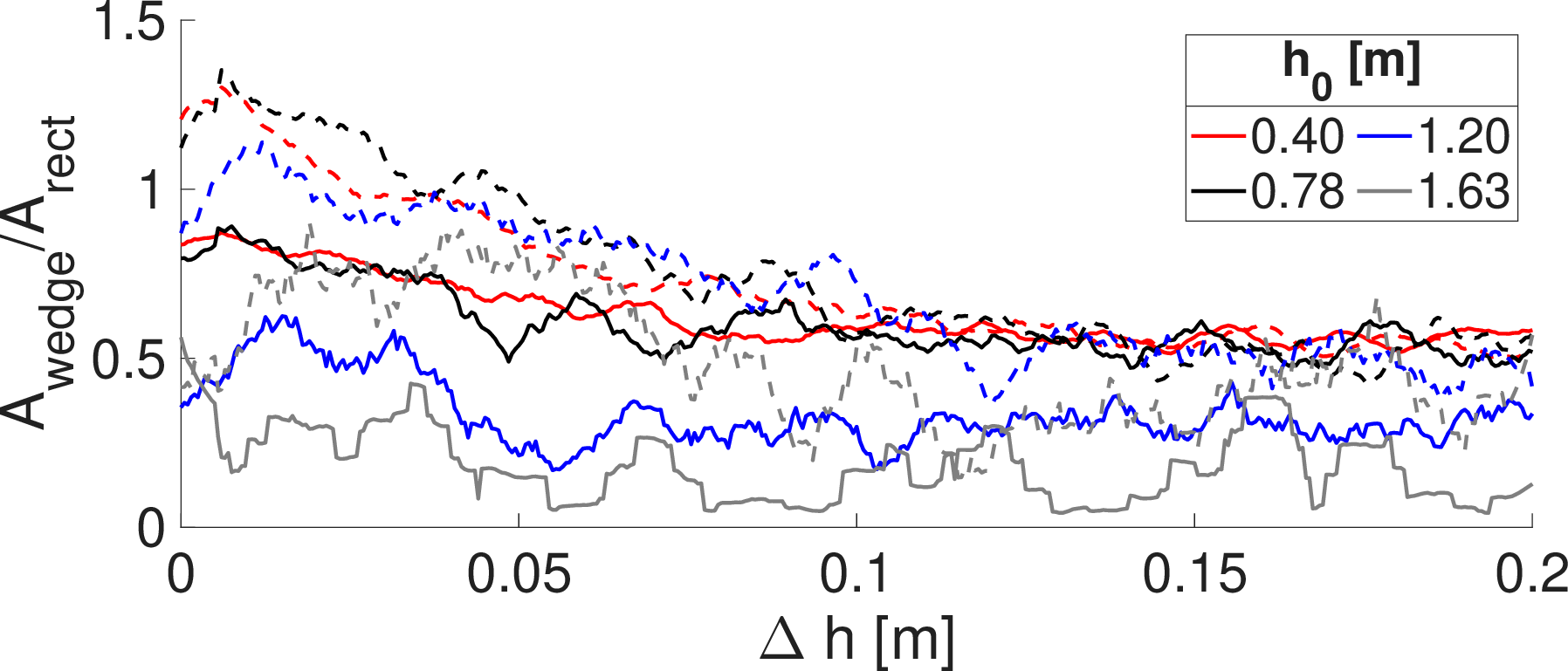}
 \caption{Evolution of the size of the uplift wedge with displacement $\Delta h$, relative to the depth of the plate, for different initial embedding depth $h_0$. Solid lines represent $\mu=0.1$, dashed lines $\mu=1.0$. The curves for $h_{0}\leq1.20$|\,m are sample averages, $h_{0}=1.63$\,m represents a single sample.}
 \label{fig:wedge_size}
\end{figure}

As Figs.\,\ref{fig:forcechain_velo_peak} and \ref{fig:forcechain_velo_steady} have shown, the shape of the wedge moving upwards with the plate depends on the initial embedding depth $h_0$, the displacement $\Delta h$ and the friction between the particles. We can measure the total affected area $A_{\mathrm{wedge}}$ by first removing all particles that move slower than 75\% of the plate velocity, and then removing all remaining particles that are not connected to the plate by a force path. The extraction of the uplift wedge is described in more detail in\,\ref{app:wedge_isolation}. To ensure comparability across different $h_{0}$ we normalize the $A_{\mathrm{wedge}}$ with the area of a rectangle stretching from the plate to the surface of the aggregate, $A_{\mathrm{rect}} = l_{\mathrm{plate}}\cdot h$. The uplift-induced deformation of the surface for $h_{0}\geq0.4\,$m is small enough to not affect the result, for $h_{0}=0.2\,$m, the plate is pulled out of the aggregate and the normalization fails. The resulting evolution of the relative area $A_{\mathrm{wedge}}/A_{\mathrm{plate}}$ is shown in Fig.\,\ref{fig:wedge_size}. 

For $h_{0}<1.0$\,m the curves are essentially independent of $h_{0}$: An initial increase in $A_{\mathrm{wedge}}/A_{\mathrm{plate}}$ is followed by a decrease after the material fails and convergence onto a constant value of $A_{\mathrm{wedge}}/A_{\mathrm{plate}}\approx 0.5$ in the steady state. Larger values of $\mu$ lead to a higher maximum and stronger initial increase in $A_{\mathrm{wedge}}/A_{\mathrm{plate}}$. The final steady-state value of $A_{\mathrm{wedge}}/A_{\mathrm{plate}}$ is independent of both $\mu$ and $h_{0}$, generalizing our previous results from\,\cite{Krengel2026} to the initial embedding depth of the plate. For $A_{\mathrm{wedge}}/A_{\mathrm{plate}}<1$ the contour of $A_{\mathrm{plate}}$ is the bounding rectangle of the uplifted volume. Assuming that the failure in the aggregate is symmetric around the plate, $A_{\mathrm{wedge}}/A_{\mathrm{plate}}=0.5$ implies that the shape of the uplifted aggregate is an isosceles triangle.

For $h_{0}\geq1.20$\,m the curve behaves differently. For large $\mu$ the initially value of $A_{\mathrm{wedge}}/A_{\mathrm{plate}}$ and its first increase decrease with $h_{0}$, but the data still converges onto the same steady state value as for $h_{0}\leq0.78$\,m. When $\mu$ is small, the mobilized particle volume in the steady state is significantly lower, instead of the triangular wedge a roughly semi-circular area is emerging, as shown also in Fig.\,\ref{fig:forcechain_velo_steady}.

\begin{figure}[h]
 \centering
 \includegraphics[width=\columnwidth]{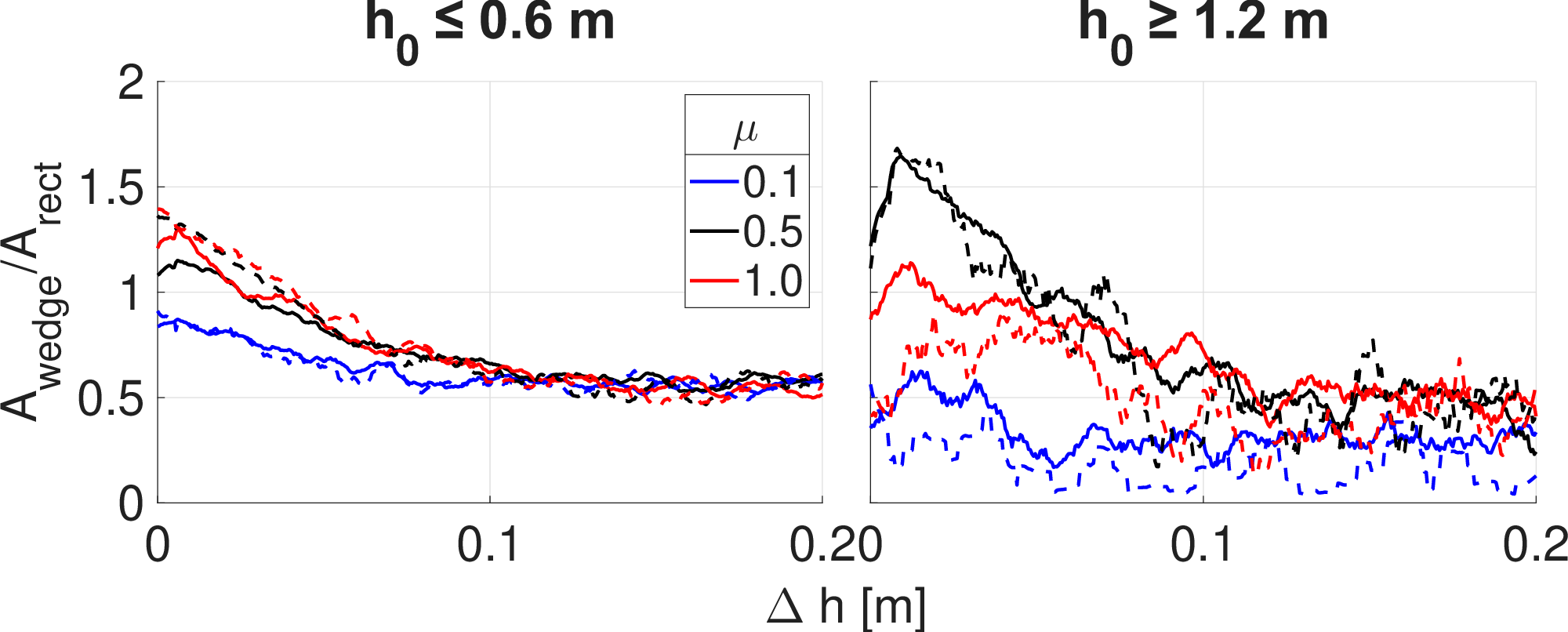}
 \caption{Correlation between $h_{0}$ and $\mu$. Dashed and solid lines represent a different initial embedding depth of the plate in each subfigure.}
 \label{fig:wedge_mu}
\end{figure}
Varying the embedding depths also changes how different aggregates resist the plate movement before failure, see Fig.\,\ref{fig:wedge_mu}. For $h_0\leq0.6$\,m the plate uplifts a larger volume if $\mu$ is larger.  However, for $h_0\geq1.2$\,m, we find that the largest volume is uplifted during peak at intermediate $\mu$. For $h_0=0.78$\,m we find a transition regime\,\cite{Fujikata2026} where the same behaviour as for deeper embedding begins to emerge, but the precise response to $\mu$ still varies with the local particle packing.

\subsection{Interaction between plate and walls}
\begin{figure}[b!]
 \centering
 \includegraphics[width=\columnwidth]{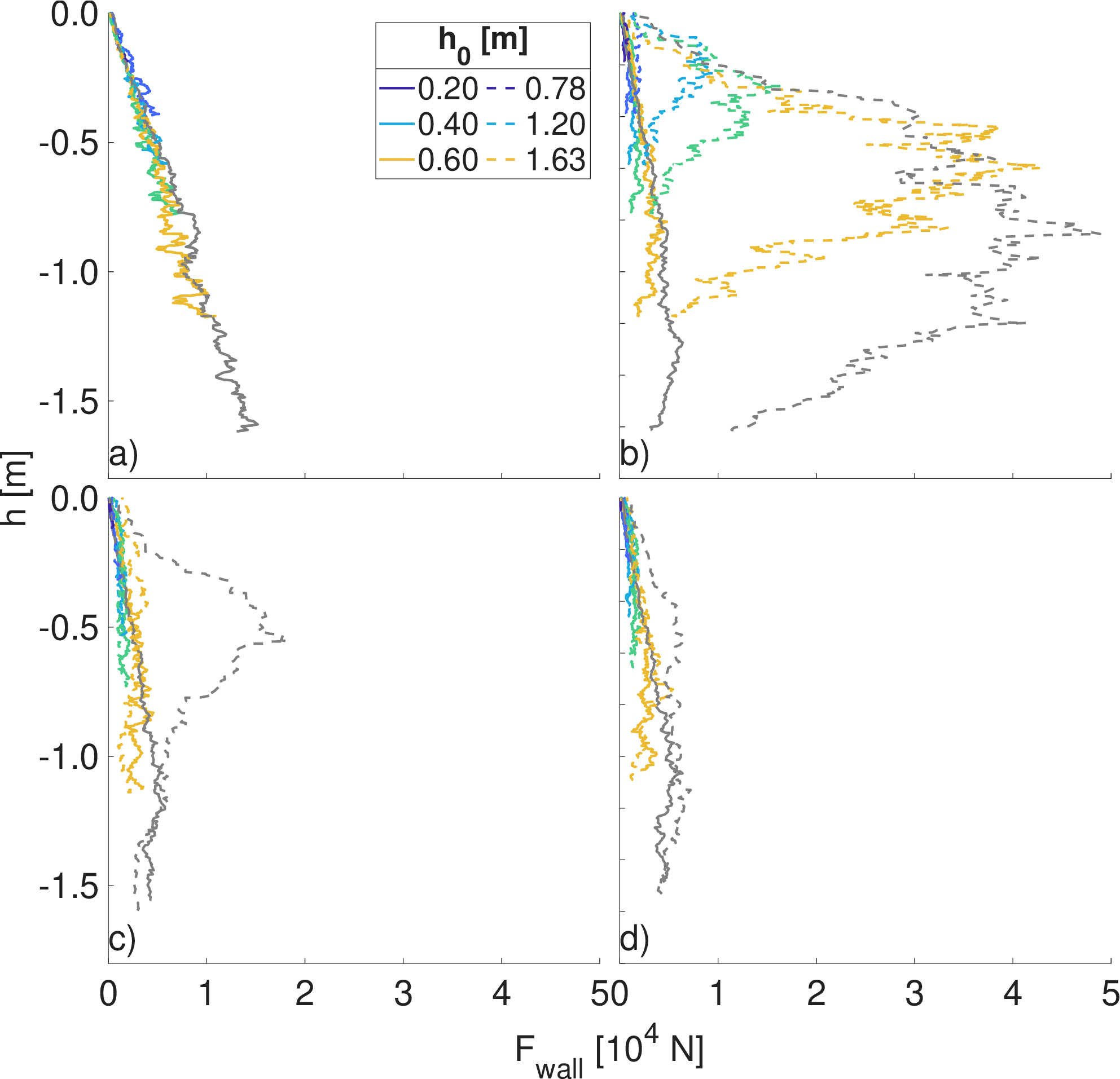}
 \caption{Distribution and magnitude of the normal forces on the sidewalls of the container with respect to the depth below the free surface, $h$: a) at the beginning of pullout, b) during peak resistance, c) at the onset of the steady state, and d) in the middle of the steady state. Solid lines mark $\mu=0.1$, dashed lines $\mu=1.0$.}
 \label{fig:wallforce}
\end{figure}
As shown in Sec.\,\ref{sec:forcechains}, for $\mu\geq0.5$ and $h_0\geq0.78$ significant interaction between the plate and the container walls occurs before failure, indicating that the walls play some role in the peak drag force of the plate. In contrast, after failure the aggregate between the plate and the walls dilates and the force chains point upwards. In this section we analyse the profile of normal forces along the container walls with depth below surface, $h$, to quantify how the plate interacts with the walls.
Figure\,\ref{fig:wallforce}, a) shows that the initial distribution of the normal forces acting on the container walls increases linear with $h$ until the top surface of the plate. As the configuration was initialized without friction, the profile is identical for all $\mu$; i.e. the influence of friction during equilibration is negligible for a dense packing. Likewise, saturation of the force due to the Janssen effect has not yet set in\,\cite{Janssen1895,Sperl2005}. Once friction is enabled, some saturation begins to show at around $h=1.3$\,m, roughly matching the predicted transition depth\,\cite{Wambaugh2010},
\begin{equation}
 \lambda = \frac{W}{2\mu K} \sim 1.25\,\mathrm{m},
\end{equation}
for our given configuration, and $K=0.4$ for polygonal particles. However, the movement of the plate quickly becomes dominant and the saturation disappears. The induced force chains emanate from the plate at a specific angle and reach the container walls at different locations above the plate (see also Fig.\,\ref{fig:forcechain_velo_peak}), which leads to a localised force maximum and a drop in force closer to the plate (Fig.\,\ref{fig:wallforce}, b). 

For each value of $\mu$, the force profile seems to follow a master curve with an approximately linear dependence on $h$ until it reaches a maximum that depends on $h_0$, both in magnitude and location. At low $\mu$ this maximum occurs closer to the plate location than at high $\mu$. For plates close to the surface saturation still is possible as the force chains no longer reach the walls but instead end at the free surface. As the aggregate fails, the force peak migrates upwards and decreases in size (Fig\,\ref{fig:wallforce},c) and d), the aggregate around the plate is too dilated now so that strong force chains no longer reach the wall. Instead, we observe that the effect of friction on the force profile begins to vanish near the plate. The dilated zone essentially acts as a limiter for force transfer into the lateral walls, so that only a part of the aggregate is influenced by wall effects.

\begin{figure}[h]
 \centering
 \includegraphics[width=\columnwidth]{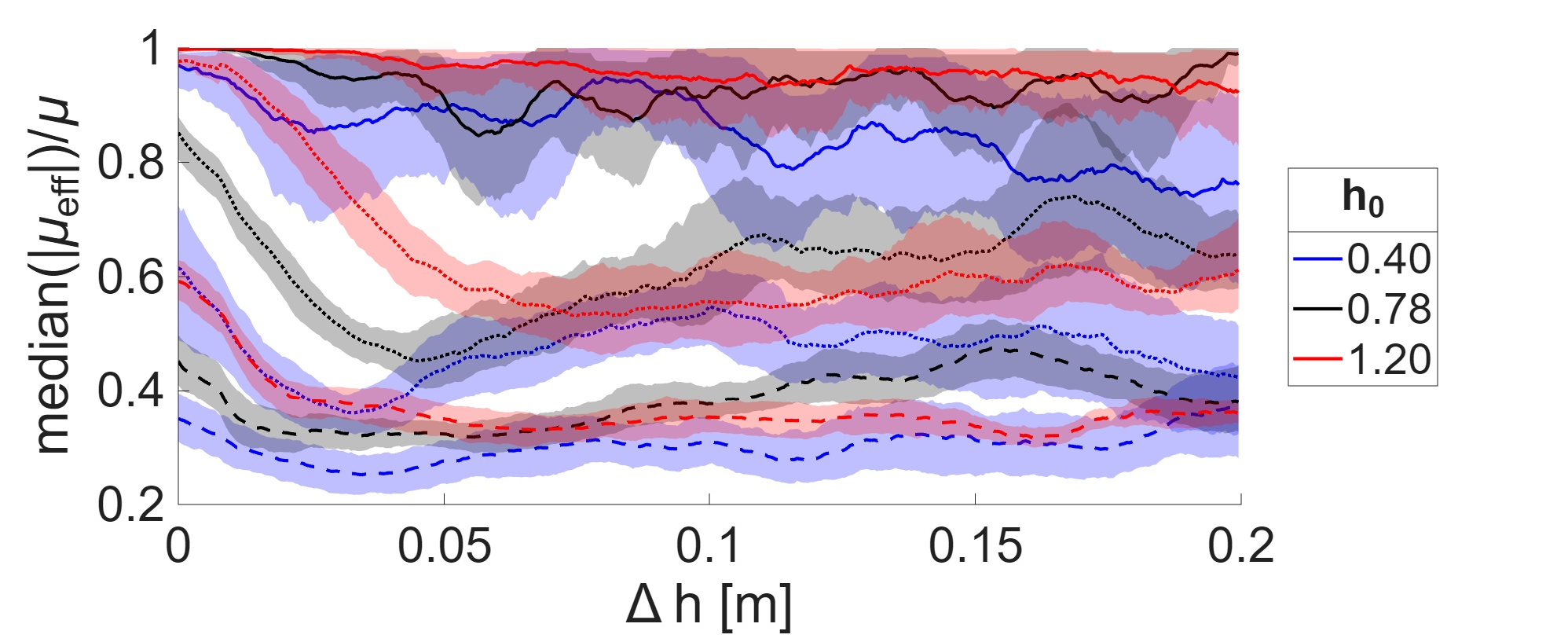}
 \caption{Evolution of the median mobilized friction magnitude median($|\mu_{\mathrm{eff}}|$)$/\mu$ on the walls, scaled with the prescribed friction coefficient, with plate displacement $\Delta h$ for different initial embedding depths $h_0$. The lines mark the average of three samples, the shadow their respective minimum and maximum values. Solid lines show $\mu=0.1$, dotted lines $\mu=0.5$ and dashed lines $\mu=1.0$.}
 \label{fig:wallfric}
\end{figure}

Frictional contacts with the container walls additionally stabilizes the aggregate. In Fig.\,\ref{fig:wallfric} we show the median value of the friction force magnitude,
\begin{equation}
 \text{median}(|\mu_{\mathrm{eff}}|)=\text{median}\left(\frac{|F_{\mathrm{T}}|}{F_{\mathrm{N}}}\right),
\end{equation}
 where $F_{\mathrm{N}}$ and $F_{\mathrm{T}}$ are the normal and tangential contact forces for all contacts along the walls. We chose to use the magnitude instead of the directed tangential contact force as the force fluctuates between positive and negative values between timesteps due to the quasi-static loading and the mean and median would thus fall around zero.
Overall, the mobilized friction across the wall primarily depends more on the applied friction coefficient $\mu$ than on the embedding depth $h_0$. Low values of $\mu=0.1$ lead to a greater mobilization of friction, $\text{median}(|\mu_{\mathrm{eff}}|)/\mu\lesssim 1$, which remains at this level throughout the plate pullout.  For large $\mu=1.0$, the mobilized friction slightly drops in the beginning as the plate begins to move, but then remains at this level. In contrast, for $\mu=0.5$ friction is nearly maximal at the beginning, followed by a steep decrease and then an increase to a stable level between the values for $\mu=0.1$ and $\mu=1.0$. In all cases, deeper embedding, i.e. greater $h_0$, leads to greater initial mobilization of friction. After dilation, around the onset of the steady state ($\Delta h=0.08$\,m), the differences between moderate ($h_0=0.78$\,m) and deep embedding ($h_0=1.20$\,m) disappear, with the mobilization for $h_0=0.78$\,m even partially being greater than for deeper embedding. Essentially, what is happening is, that for sufficiently large friction (here $\mu\geq0.5$) the aggregate dilates so much that the force chains emanating from the plate can no longer reach the walls. The interaction forces between the aggregate and the walls saturate, and any further increase in $\mu$ leads to the same mobilization in friction.

\section{Conclusions}
In this work we investigated the constant-velocity pullout of a plate-like intruder from a two-dimensional, dense granular medium at different initial embedding depths at a fixed system size.

The first finding is that the maximum drag force shows different scaling with embedding depth $h_0$ depending on the friction coefficient $\mu$, where the drag force is independent of $\mu$ at near surface embedding, but depends linear on $h_0$ if $\mu$ is small, and non-linear if $\mu$ is large. A similar scaling is found for the steady-state drag force, however, the transition depth at which $\mu$ takes effect differs between the steady state and the peak force measurements.

The second finding is, that the different scaling of the peak drag force is due to the influence of the container walls becoming notable at larger values of $\mu$, which amplifies the resistance to the plate motion. Regardless of the embedding depth, if $\mu$ is sufficiently large the plate induces large forces on the domain boundaries that follow the same depth-force profile. In the steady state instead the interaction between walls and the plate becomes minimal again for all cases, as the area between the plate and the walls has dilated and forces are directed upwards instead of outwards.

The third finding is, that the steady state instead shows that for weakly frictional materials failure depends on the embedding depth: After a certain critical depth the overhead pressure is large enough to overcome the frictional resistance between particles surrounding the plate, so no notable dilation occurs, and the failure is limited to a semi-spherical failure line on top of the plate. Around the plate edges, dilation still occurs to allow particle transport to below the plate, but above the plate, the aggregate remains unchanged. 

Our study here was conducted with a fixed size-ratio between the plate and the granular domain. Friction controls how far force chains can extend. This means, for a larger container, the influence of the container walls will become notable only at higher friction coefficients and the peak and steady state drag forces may also scale linear at larger $\mu$. On the other hand, friction also affects the competition between sliding and rolling, and for sufficiently large $\mu$ sliding is essentially prohibited. We expect, that in this case, the drag force will also scale differently, even in the absence of walls, however a proof of our conjecture is still outstanding.


\section*{CRediT authorship contribution statement}
\textbf{D. Krengel:} Writing - original draft, Writing - review \& editing, Software, Methodology, Validation, Investigation, Formal analysis, Data curation, Conceptualization, Visualisation, Supervision. \\
\textbf{J. Chen:} Writing - review \& editing, Software, Data curation.\\
\textbf{S. Nomura:} Writing - review \& editing, Resources.\\
\textbf{H. Takahashi:} Writing - review \& editing.

\section*{Declaration of competing interests}
The authors declare that they have no known competing financial interests or personal relationships that could have appeared to influence the work reported in this paper.
\section*{Data availability}
Data is available from the corresponding author upon reasonable request.
\section*{Acknowledgments}
The authors are grateful for the support by the Grant (JPMJFR205U) from the Japan Science and Technology Agency (JST) and for the Grant-in-Aid for Scientific Research (22K04306) from the Japan Society for the Promotion of Science (JSPS). This study was supported by the Earth Simulator project of the Japan Agency for Marine-Earth Science and Technology (JAMSTEC).

\appendix
\section{Polygonal Discrete Element Method}\label{app:DEM}
We use a two-dimensional `hard-particle soft-contact' DEM\,\cite{Matuttis2014}. Here, particles are modelled as convex polygons instead of discs to capture the effect of shape on the granular micromechanics. 
\begin{figure}[b!]
  \centering
  \includegraphics[width=0.8\columnwidth]{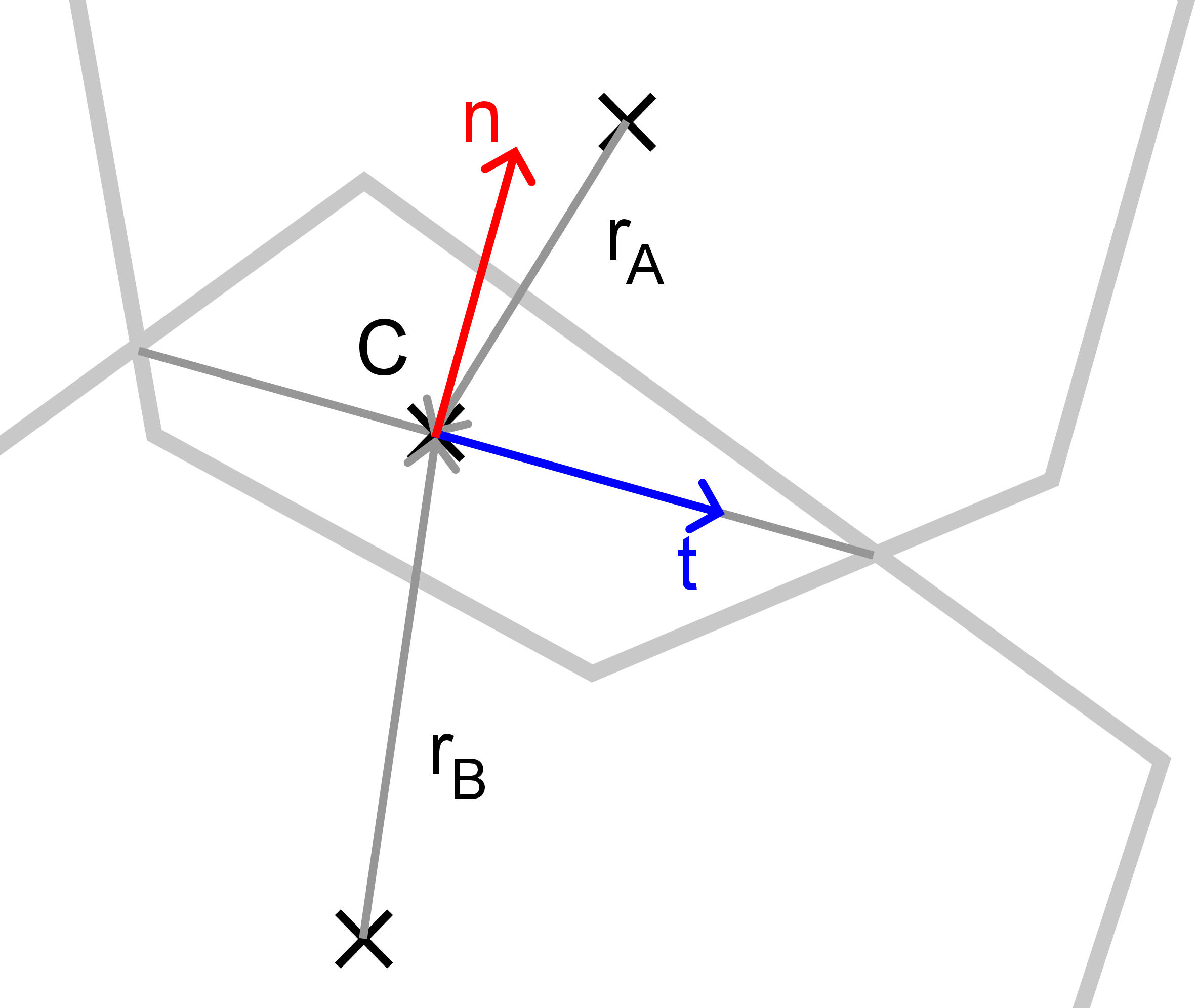}
  \caption{Geometry of the contact between two polygonal particles. For the sake of visualisation the scale of the overlap is significantly exaggerated.}
  \label{fig:DEM_method}
\end{figure}

Figure\,\ref{fig:DEM_method} shows the contact geometry for a particle pair.
For two contacting polygons $A$ and $B$ in contact, the intersection line defines the tangential direction $\mathbf{t}$. This also fixes the normal direction $\mathbf{n}$. The elastic force
\begin{equation}
 F_{\mathrm{el}}=\frac{Y\cdot A}{l}
\end{equation}
is proportional to the overlap area $A$ and the Young's modulus $Y$. A characteristic length
\begin{equation}
 l=4\frac{|\mathbf{r}_{A}|\cdot|\mathbf{r}_{B}|}{|\mathbf{r}_{A}|+|\mathbf{r}_{B}|},
\end{equation}
based on the contact vectors $\mathbf{r}_{A,B}$ between the particle centres of mass and the centroid $\textbf{C}$ of the overlap area, is required to define the normal force in units of [N]. Further, a dissipative force
\begin{equation}
 F_{\mathrm{diss}}=\gamma \sqrt{mY}\frac{\dot{A}}{l},
\end{equation}
with the reduced mass $1/m = 1/m_{A} + 1/m_{B}$ and a damping constant $\gamma$, is acting in normal direction, so that the total normal force is
\begin{equation}
 F_{N} = F_{\mathrm{el}}+F_{\mathrm{diss}}.
\end{equation}
A friction force in tangential direction is implemented through the Cundall-Strack model\,\cite{Cundall1979},
\begin{equation}
 F_{\mathrm{T}}(t) = \left\{\begin{array}{ll}
        F_{\mathrm{T}}(t-\mathrm{d}t)-k_{\mathrm{t}}v_{\mathrm{t}}\mathrm{d}t & |F_{\mathrm{T}}(t)|\leq \mu F_{\mathrm{N}}\\
				\mathrm{sgn}\left(F_{\mathrm{T}}(t-\mathrm{d}t)\right)\mu F_{\mathrm{N}} & |F_{\mathrm{T}}(t)| > \mu F_{\mathrm{N}}
\end{array}\right.,
\end{equation}
where $v_{\mathrm{t}}$ is the relative tangential velocity at the contact point and $k_{\mathrm{t}}$ the `tangential stiffness' of a spring. The total force $\mathbf{F} = F_{\mathrm{N}}\mathbf{n} + F_{\mathrm{T}}\mathbf{t}$ acts upon the centroid of the overlap area which induces a torque on each particle as
\begin{equation}
 \bm{\tau}_{\mathrm{A,B}} = \mathbf{r}_{\mathrm{A,B}}\times\mathbf{F}.
\end{equation}

\section{Isolation of the uplift wedge}\label{app:wedge_isolation}
\begin{figure}[h!]
 \centering
 \includegraphics[width=0.475\columnwidth]{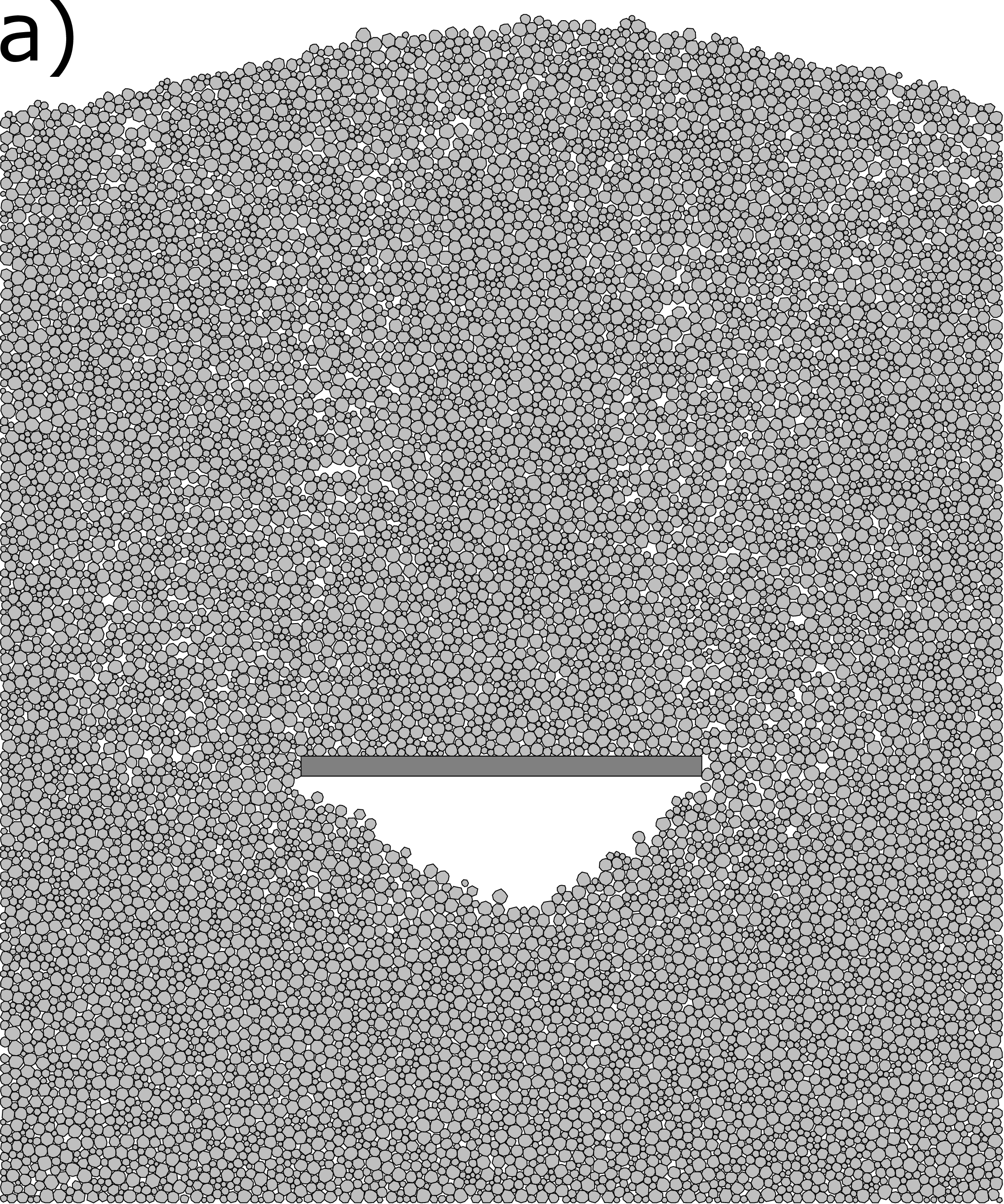}
 \hfill
 \includegraphics[width=0.475\columnwidth]{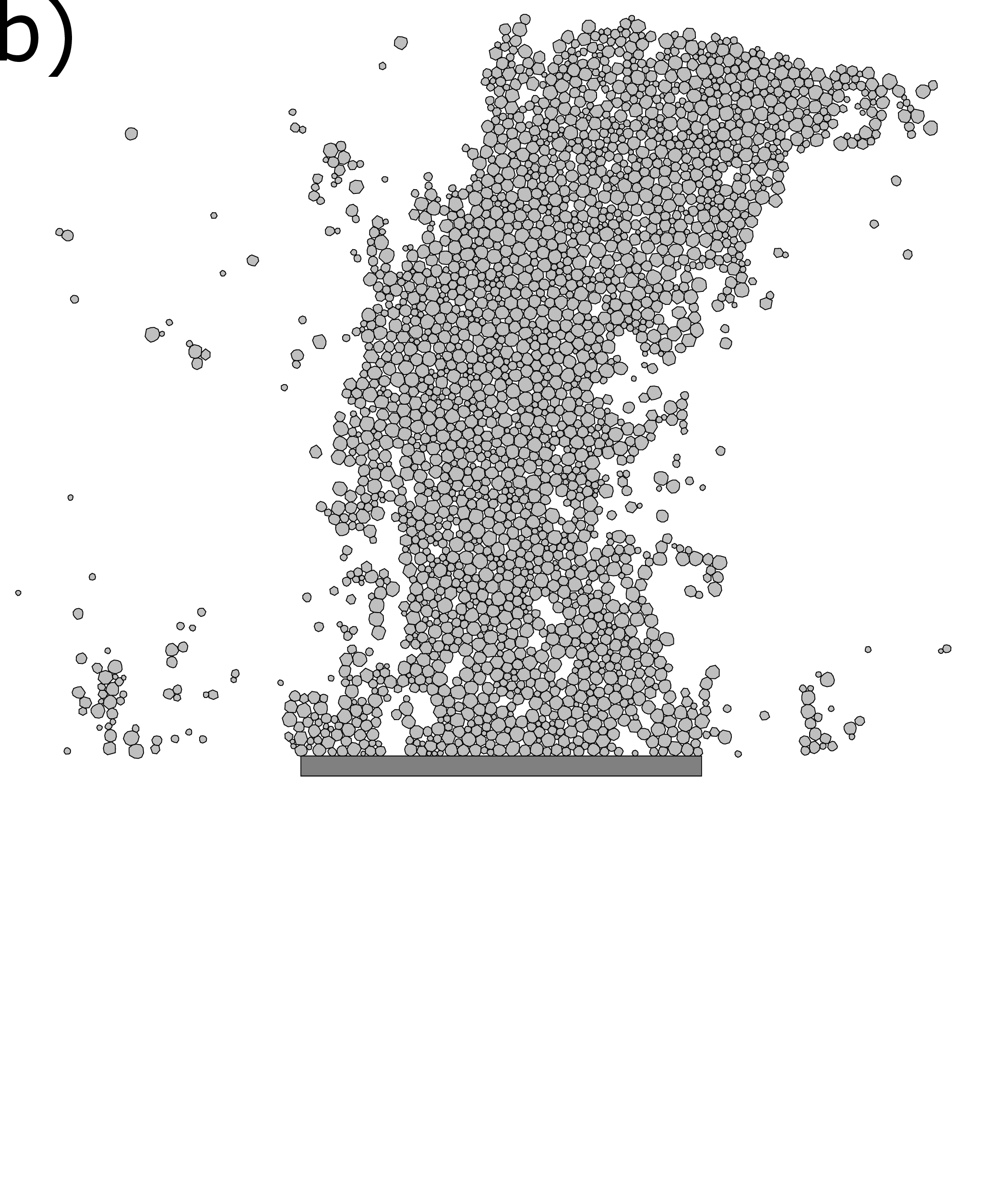}
 
 \includegraphics[width=0.475\columnwidth]{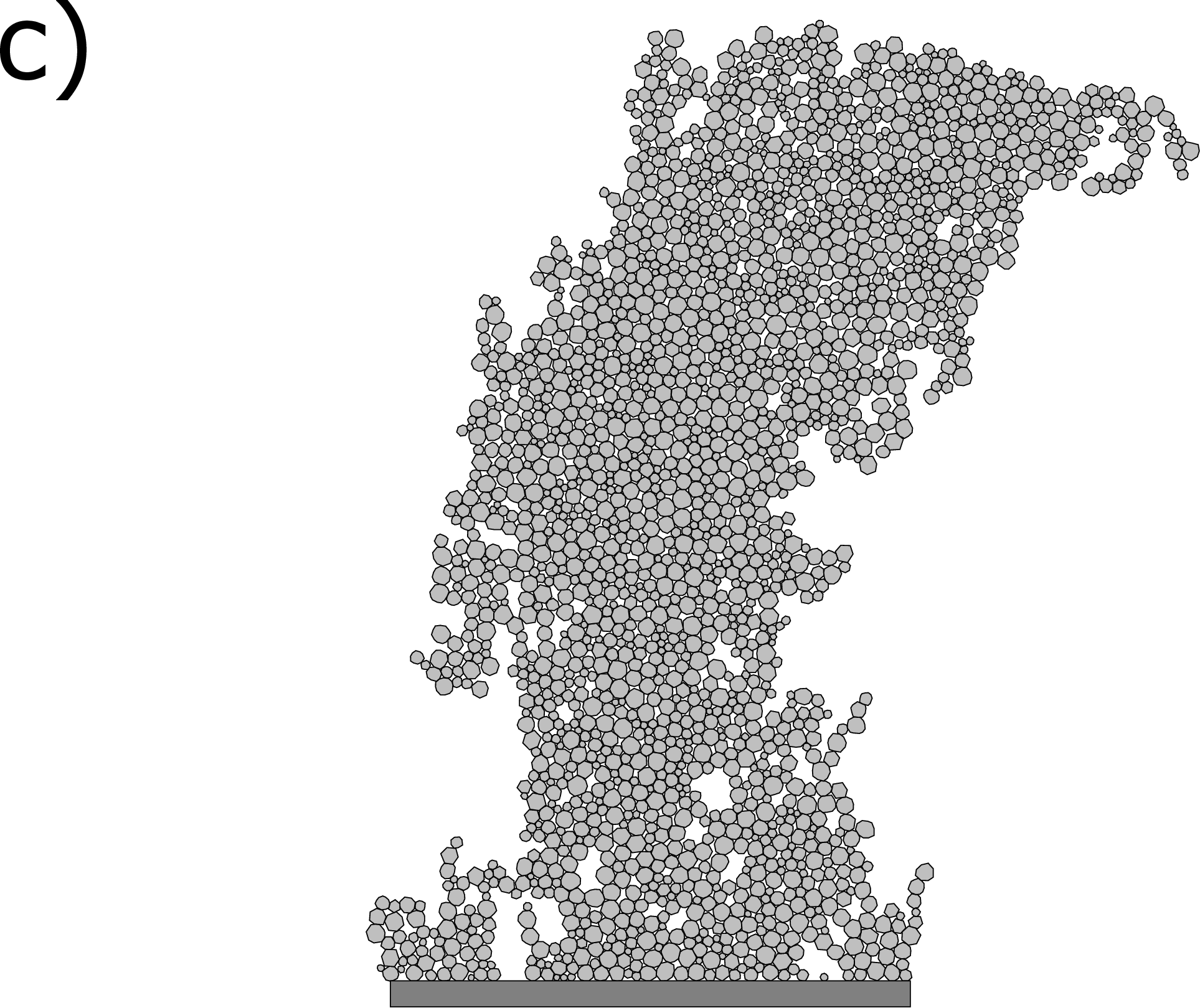}
 \hfill
 \includegraphics[width=0.475\columnwidth]{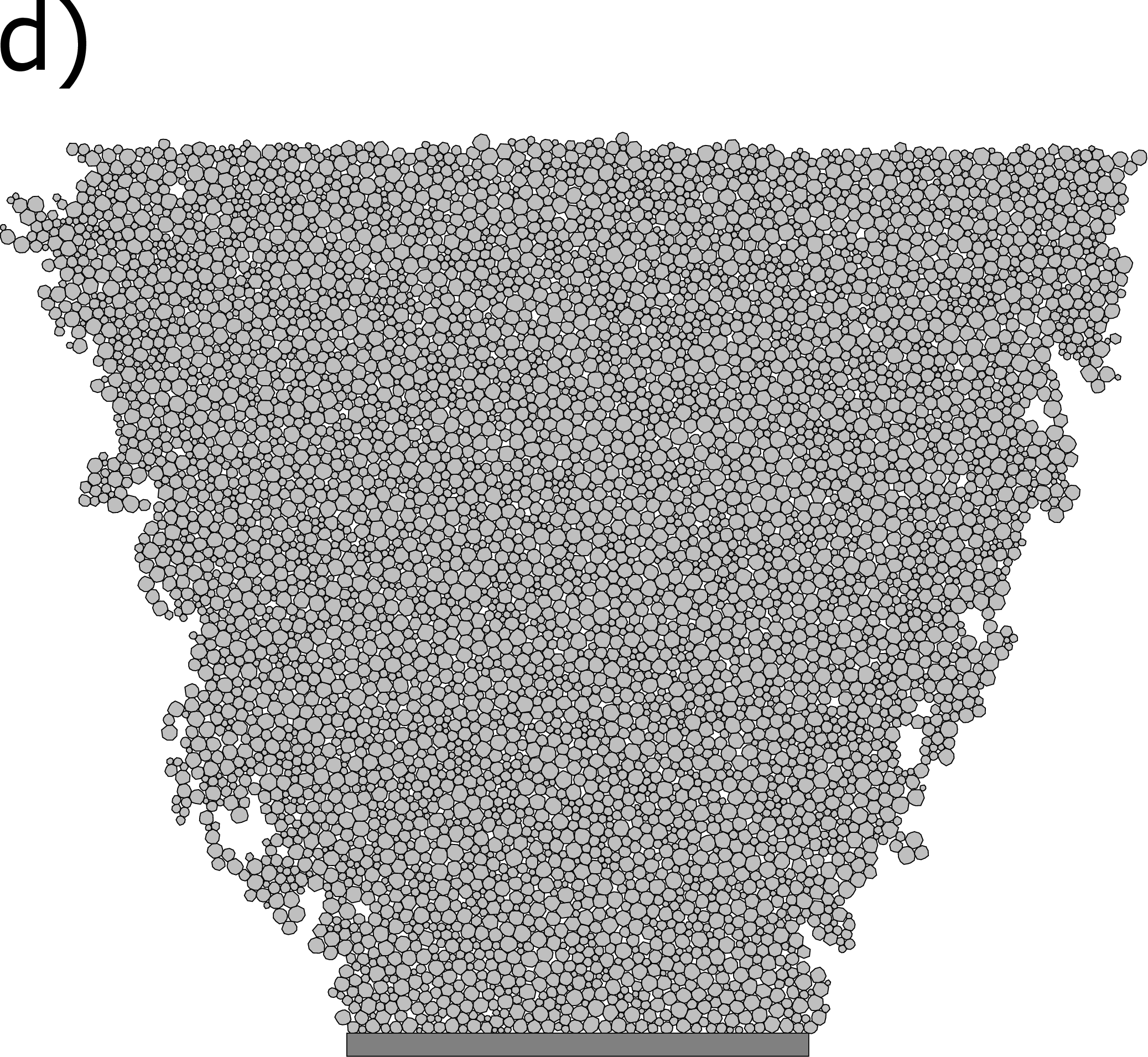}
 \caption{Filtering of the a) initial particle configuration by b) vertical velocity and position relative to the plate and c) by force path connections to the plate. a)-c) show the configuration at the onset of the ``steady state'' ($\Delta h = 0.08$\,m), and d) shows the isolated uplift wedge during maximum drag force.}
 \label{fig:isolate_wedge}
\end{figure}

In this section we detail our method to isolate the uplift wedge. At first, each particle with a vertical velocity $v_{\mathrm{particle}, y}< 0.8 v_{\mathrm{plate}, y}$, or that is below the upper plate edge, is removed from the original data set (Fig.\,\ref{fig:isolate_wedge}, a) to isolate only the particles that move upwards (Fig.\,\ref{fig:isolate_wedge} b). Then we remove any remaining particle that has no force path connecting it to the intruder (Fig.\,\ref{fig:isolate_wedge} c). The resulting structure is the uplifted material of the system. Until failure, it will generally be an outward pointing trapezoid (Fig.\,\ref{fig:isolate_wedge} d), after failure it will typically converge onto an isosceles triangle form. While the final shape can strongly fluctuate, its time average is generally stable.  The threshold velocity is chosen less than the plate velocity to allow for momentarily velocity fluctuations of the individual particles. The force path condition typically removes any erroneously kept particles. The threshold value can be adjusted, but in our experience, the structures resulting from $0.8 v_{\mathrm{plate}, y}$ most reliable matched the visualisation of the uplift wedge shown in e.g. Fig.\,\ref{fig:forcechain_velo_peak}.
The benefit of this method is that it is easily automatable still provides a qualitative understanding how the shape of the uplift wedge evolves during the pullout process. Using a cumulative rotation filter would allow to isolate the failure surface much clearer than filtering by vertical velocity, but only works  for sufficiently large displacement when enough rotation has accumulated, and thus cannot be used in the earlier stages of the pullout process.

\bibliographystyle{elsarticle-num} 
\bibliography{intruder_depth}

@Article{Krengel2026,
  author    = {Krengel, Dominik and Chen, Jian and Nomura, Shun and Nishiura, Daisuke and Takahashi, Hidenori},
  journal   = {Ocean Engineering},
  title     = {The influence of plate shape on the mechanics of an expandable plate anchor in cohesionless soil},
  year      = {2026},
  issn      = {0029-8018},
  month     = Oct,
  pages     = {127738},
  volume    = {366},
  doi       = {10.1016/j.oceaneng.2026.127738},
  publisher = {Elsevier BV},
}

@Article{Fujikata2026,
  author  = {Rikuto Fujikata and Shun Nomura and Hidenori Takahashi},
  journal = {manuscript in preparation},
  title   = {Holding capacity and failure mode transition during pullout of plate anchors in sand from shallow to deep},
  year    = {2026},
}

@Article{Hossain2020,
  author    = {Hossain, T. and Rognon, P.},
  journal   = {Granular Matter},
  title     = {Rate-dependent drag instability in granular materials},
  year      = {2020},
  issn      = {1434-7636},
  month     = Aug,
  number    = {3},
  volume    = {22},
  doi       = {10.1007/s10035-020-01039-5},
  publisher = {Springer Science and Business Media LLC},
}

@Article{Cundall1979,
  author    = {P. A. Cundall and O. D. L. Strack},
  journal   = {G{\'{e}}otechnique},
  title     = {A discrete numerical model for granular assemblies},
  year      = {1979},
  month     = {mar},
  number    = {1},
  pages     = {47--65},
  volume    = {29},
  doi       = {10.1680/geot.1979.29.1.47},
  publisher = {Thomas Telford Ltd.},
}

@Book{Matuttis2014,
  author    = {{Matuttis, H.-G.} and Chen, J.},
  publisher = {Wiley},
  title     = {Understanding the {D}iscrete {E}lement {M}ethod: {S}imulation of Non-Spherical Particles for Granular and Multi-body Systems},
  year      = {2014},
  address   = {Singapore},
  edition   = {1},
}

@Article{Sperl2005,
  author    = {Sperl, Matthias},
  journal   = {Granular Matter},
  title     = {Experiments on corn pressure in silo cells – translation and comment of Janssen’s paper from 1895},
  year      = {2005},
  issn      = {1434-7636},
  month     = Dec,
  number    = {2},
  pages     = {59--65},
  volume    = {8},
  doi       = {10.1007/s10035-005-0224-z},
  publisher = {Springer Science and Business Media LLC},
}

@Article{Janssen1895,
  author  = {Janssen, H. A.},
  journal = {Zeitschrift des Vereines deutscher Ingenieure},
  title   = {Versuche über {G}etreidedruck in {S}ilozellen},
  year    = {1895},
  pages   = {1045-1049.},
  volume  = {39},
}

@Article{Boonkanokwong2016,
  author    = {Boonkanokwong, Veerakiet and Remy, Brenda and Khinast, Johannes G. and Glasser, Benjamin J.},
  journal   = {Powder Technology},
  title     = {The effect of the number of impeller blades on granular flow in a bladed mixer},
  year      = {2016},
  issn      = {0032-5910},
  month     = Nov,
  pages     = {333--349},
  volume    = {302},
  doi       = {10.1016/j.powtec.2016.08.064},
  publisher = {Elsevier BV},
}

@Article{Nadeem2021,
  author    = {Nadeem, Humair and Subramaniam, Shankar and Sinha, Kushal and Heindel, Theodore J.},
  journal   = {Powder Technology},
  title     = {Characterization of intruder particle motion in a bladed mixer},
  year      = {2021},
  issn      = {0032-5910},
  month     = Mar,
  pages     = {440--450},
  volume    = {381},
  doi       = {10.1016/j.powtec.2020.12.002},
  publisher = {Elsevier BV},
}

@Article{Nomura2026,
  author    = {Nomura, Shun and Sato, Arata and Tani, Kazuo},
  journal   = {Soils and Foundations},
  title     = {Concept of Expandable Plate Anchor (ExPLA) for floating offshore wind and its evaluation through 1 g model tests},
  year      = {2026},
  issn      = {0038-0806},
  month     = Oct,
  number    = {5},
  pages     = {101830},
  volume    = {66},
  doi       = {10.1016/j.sandf.2026.101830},
  publisher = {Elsevier BV},
}

@Article{Lehuen2020,
  author    = {Lehuen, Julien and Delenne, Jean-Yves and Duri, Agnès and Ruiz, Thierry},
  journal   = {Granular Matter},
  title     = {Forces and flow induced by a moving intruder in a granular packing: coarse-graining and DEM simulations versus experiments},
  year      = {2020},
  issn      = {1434-7636},
  month     = sep,
  number    = {4},
  pages     = {-},
  volume    = {22},
  doi       = {10.1007/s10035-020-01047-5},
  publisher = {Springer Science and Business Media LLC},
}

@Article{Radl2012,
  author    = {Radl, Stefan and Brandl, Daniel and Heimburg, Hanna and Glasser, Benjamin J. and Khinast, Johannes G.},
  journal   = {Powder Technology},
  title     = {Flow and mixing of granular material over a single blade},
  year      = {2012},
  issn      = {0032-5910},
  month     = aug,
  pages     = {199--212},
  volume    = {226},
  doi       = {10.1016/j.powtec.2012.04.042},
  publisher = {Elsevier BV},
}

@Article{Kurniadi2025,
  author    = {Kurniadi, R. and Roy, A. and Maitra, S. and Chow, S.H. and Cassidy, M.J.},
  journal   = {Computers and Geotechnics},
  title     = {Uplift mechanism of horizontal circular plate anchors under varying drainage conditions in sand},
  year      = {2025},
  issn      = {0266-352X},
  month     = apr,
  pages     = {107062},
  volume    = {180},
  doi       = {10.1016/j.compgeo.2025.107062},
  publisher = {Elsevier BV},
}

@Article{Liang2021,
  author    = {Liang, Weijian and Zhao, Jidong and Wu, Huanran and Soga, Kenichi},
  journal   = {Journal of Geotechnical and Geoenvironmental Engineering},
  title     = {Multiscale Modeling of Anchor Pullout in Sand},
  year      = {2021},
  issn      = {1943-5606},
  month     = sep,
  number    = {9},
  pages     = {-},
  volume    = {147},
  doi       = {10.1061/(asce)gt.1943-5606.0002599},
  publisher = {American Society of Civil Engineers (ASCE)},
}

@Article{Liu2012,
  author    = {Liu, Jinyuan and Liu, Mingliang and Zhu, Zhende},
  journal   = {Journal of Geotechnical and Geoenvironmental Engineering},
  title     = {Sand Deformation around an Uplift Plate Anchor},
  year      = {2012},
  issn      = {1943-5606},
  month     = jun,
  number    = {6},
  pages     = {728--737},
  volume    = {138},
  doi       = {10.1061/(asce)gt.1943-5606.0000633},
  publisher = {American Society of Civil Engineers (ASCE)},
}

@Article{Sivakumar2013,
  author    = {Sivakumar, V. and O’Kelly, B.C. and Madhav, M.R. and Moorhead, C. and Rankin, B.},
  journal   = {Canadian Geotechnical Journal},
  title     = {Granular anchors under vertical loading – axial pull},
  year      = {2013},
  issn      = {1208-6010},
  month     = feb,
  number    = {2},
  pages     = {123--132},
  volume    = {50},
  doi       = {10.1139/cgj-2012-0203},
  publisher = {Canadian Science Publishing},
}

@Article{Krengel2026a,
  author    = {Krengel, Dominik and Chen, Jian and Nomura, Shun and Ota, Shunsuke and Takahashi, Hidenori},
  journal   = {Powder Technology},
  title     = {Vertical pullout of a plate-like intruder from a granular medium: From system-wide response to avalanching around the intruder},
  year      = {2026},
  issn      = {0032-5910},
  month     = June,
  pages     = {122440},
  volume    = {476},
  doi       = {10.1016/j.powtec.2026.122440},
  publisher = {Elsevier BV},
}

@Article{Albert1999,
  author    = {Albert, R. and Pfeifer, M. A. and Barabási, A.-L. and Schiffer, P.},
  journal   = {Physical Review Letters},
  title     = {Slow Drag in a Granular Medium},
  year      = {1999},
  issn      = {1079-7114},
  month     = jan,
  number    = {1},
  pages     = {205--208},
  volume    = {82},
  doi       = {10.1103/physrevlett.82.205},
  publisher = {American Physical Society (APS)},
}

@Article{Albert2001,
  author    = {Albert, I. and Sample, J. G. and Morss, A. J. and Rajagopalan, S. and Barabási, A.-L. and Schiffer, P.},
  journal   = {Physical Review E},
  title     = {Granular drag on a discrete object: Shape effects on jamming},
  year      = {2001},
  issn      = {1095-3787},
  month     = nov,
  number    = {6},
  pages     = {061303},
  volume    = {64},
  doi       = {10.1103/physreve.64.061303},
  publisher = {American Physical Society (APS)},
}

@Article{He2025,
  author    = {He, Hantao and Zhang, Qiong and Ottino, Julio M. and Umbanhowar, Paul B. and Lueptow, Richard M.},
  journal   = {Journal of Fluid Mechanics},
  title     = {Lift and drag forces on a moving intruder in granular shear flow},
  year      = {2025},
  issn      = {1469-7645},
  month     = mar,
  pages     = {-},
  volume    = {1008},
  doi       = {10.1017/jfm.2025.87},
  publisher = {Cambridge University Press (CUP)},
}

@Article{Henann2014,
  author    = {Henann, David L. and Kamrin, Ken},
  journal   = {Physical Review Letters},
  title     = {Continuum Modeling of Secondary Rheology in Dense Granular Materials},
  year      = {2014},
  issn      = {1079-7114},
  month     = oct,
  number    = {17},
  pages     = {178001},
  volume    = {113},
  doi       = {10.1103/physrevlett.113.178001},
  publisher = {American Physical Society (APS)},
}

@Article{Seguin2016,
  author    = {Seguin, A. and Coulais, C. and Martinez, F. and Bertho, Y. and Gondret, P.},
  journal   = {Physical Review E},
  title     = {Local rheological measurements in the granular flow around an intruder},
  year      = {2016},
  issn      = {2470-0053},
  month     = jan,
  number    = {1},
  pages     = {012904},
  volume    = {93},
  doi       = {10.1103/physreve.93.012904},
  publisher = {American Physical Society (APS)},
}

@Article{Takada2019,
  author    = {Takada, Satoshi and Hayakawa, Hisao},
  journal   = {Granular Matter},
  title     = {Drag acting on an intruder in a three-dimensional granular environment},
  year      = {2019},
  issn      = {1434-7636},
  month     = nov,
  number    = {1},
  pages     = {-},
  volume    = {22},
  doi       = {10.1007/s10035-019-0973-8},
  publisher = {Springer Science and Business Media LLC},
}

@Book{Das2013,
  author    = {Das, Braja and Sanjay K. Shukla},
  publisher = {J. Ross Publishing, Incorporated},
  title     = {Earth Anchors},
  year      = {2013},
  address   = {Chicago},
  edition   = {2nd ed.},
  isbn      = {9781604277395},
  note      = {Description based on publisher supplied metadata and other sources.},
  pagetotal = {1345},
  ppn_gvk   = {1929944837},
}

@Article{Hossain2020a,
  author    = {Hossain, Tanvir and Rognon, Pierre},
  journal   = {Physical Review Fluids},
  title     = {Drag force in immersed granular materials},
  year      = {2020},
  issn      = {2469-990X},
  month     = may,
  number    = {5},
  pages     = {054306},
  volume    = {5},
  doi       = {10.1103/physrevfluids.5.054306},
  publisher = {American Physical Society (APS)},
}

@Article{Jalali2021,
  author    = {Jalali, Payman and Zhao, Yuchen and Socolar, Joshua E. S.},
  journal   = {Soft Matter},
  title     = {Static and dynamic features of granular material failure due to upward pulling of a buried sphere by a slowly increasing force},
  year      = {2021},
  issn      = {1744-6848},
  number    = {10},
  pages     = {2832--2839},
  volume    = {17},
  doi       = {10.1039/d0sm01914c},
  publisher = {Royal Society of Chemistry (RSC)},
}

@Article{Shah2020,
  author    = {Shah, Srujal and Cheng, Chao and Jalali, Payman and Kondic, Lou},
  journal   = {Soft Matter},
  title     = {Failure of confined granular media due to pullout of an intruder: from force networks to a system wide response},
  year      = {2020},
  issn      = {1744-6848},
  number    = {33},
  pages     = {7685--7695},
  volume    = {16},
  doi       = {10.1039/d0sm00911c},
  publisher = {Royal Society of Chemistry (RSC)},
}

@Article{Zhang2017,
  author    = {Zhang, Yue and Behringer, Robert},
  journal   = {EPJ Web of Conferences},
  title     = {Pulling an intruder from a granular material: a novel depinning experiment},
  year      = {2017},
  issn      = {2100-014X},
  pages     = {03040},
  volume    = {140},
  doi       = {10.1051/epjconf/201714003040},
  editor    = {Radjai, F. and Nezamabadi, S. and Luding, S. and Delenne, J.Y.},
  publisher = {EDP Sciences},
}

@Article{Murray1987,
  author    = {Murray, E. J. and Geddes, James D.},
  journal   = {Journal of Geotechnical Engineering},
  title     = {Uplift of Anchor Plates in Sand},
  year      = {1987},
  issn      = {1944-8368},
  month     = mar,
  number    = {3},
  pages     = {202--215},
  volume    = {113},
  doi       = {10.1061/(asce)0733-9410(1987)113:3(202)},
  publisher = {American Society of Civil Engineers (ASCE)},
}

@Article{Tagaya1988,
  author    = {Tagaya, Kozo and Scott, Ronald F. and Aboshi, Hisao},
  journal   = {Soils and Foundations},
  title     = {Pullout Resistance of Buried Anchor in Sand},
  year      = {1988},
  issn      = {0038-0806},
  month     = sep,
  number    = {3},
  pages     = {114--130},
  volume    = {28},
  doi       = {10.3208/sandf1972.28.3_114},
  publisher = {Elsevier BV},
}

@Article{Trautmann1985,
  author    = {Trautmann, Charles H. and O’Rourfce, Thomas D. and Kulhawy, Fred H.},
  journal   = {Journal of Geotechnical Engineering},
  title     = {Uplift Force‐Displacement Response of Buried Pipe},
  year      = {1985},
  issn      = {1944-8368},
  month     = sep,
  number    = {9},
  pages     = {1061--1076},
  volume    = {111},
  doi       = {10.1061/(asce)0733-9410(1985)111:9(1061)},
  publisher = {American Society of Civil Engineers (ASCE)},
}

@Article{White2008,
  author    = {White, D. J. and Cheuk, C. Y. and Bolton, M. D.},
  journal   = {Géotechnique},
  title     = {The uplift resistance of pipes and plate anchors buried in sand},
  year      = {2008},
  issn      = {1751-7656},
  month     = dec,
  number    = {10},
  pages     = {771--779},
  volume    = {58},
  doi       = {10.1680/geot.2008.3692},
  publisher = {Emerald},
}

@Article{Evans2019,
  author    = {Evans, T. Matthew and Zhang, Nan},
  journal   = {International Journal of Geomechanics},
  title     = {Three-Dimensional Simulations of Plate Anchor Pullout in Granular Materials},
  year      = {2019},
  issn      = {1943-5622},
  month     = apr,
  number    = {4},
  pages     = {-},
  volume    = {19},
  doi       = {10.1061/(asce)gm.1943-5622.0001367},
  publisher = {American Society of Civil Engineers (ASCE)},
}

@Article{Askari2016,
  author    = {Askari, Hesam and Kamrin, Ken},
  journal   = {Nature Materials},
  title     = {Intrusion rheology in grains and other flowable materials},
  year      = {2016},
  issn      = {1476-4660},
  month     = aug,
  number    = {12},
  pages     = {1274--1279},
  volume    = {15},
  doi       = {10.1038/nmat4727},
  publisher = {Springer Science and Business Media LLC},
}

@Article{Khatri2011,
  author    = {Khatri, Vishwas N. and Kumar, Jyant},
  journal   = {Canadian Geotechnical Journal},
  title     = {Effect of anchor width on pullout capacity of strip anchors in sand},
  year      = {2011},
  issn      = {1208-6010},
  month     = mar,
  number    = {3},
  pages     = {511--517},
  volume    = {48},
  doi       = {10.1139/t10-082},
  publisher = {Canadian Science Publishing},
}

@Article{Athani2017,
  author    = {Athani, S. and Kharel, P. and Airey, D. and Rognon, P.},
  journal   = {Géotechnique Letters},
  title     = {Grain-size effect on uplift capacity of plate anchors in coarse granular soils},
  year      = {2017},
  issn      = {2045-2543},
  month     = jun,
  number    = {2},
  pages     = {167--173},
  volume    = {7},
  doi       = {10.1680/jgele.17.00002},
  publisher = {Emerald},
}

@Article{Costantino2008,
  author    = {Costantino, D. J. and Scheidemantel, T. J. and Stone, M. B. and Conger, C. and Klein, K. and Lohr, M. and Modig, Z. and Schiffer, P.},
  journal   = {Physical Review Letters},
  title     = {Starting to Move through a Granular Medium},
  year      = {2008},
  issn      = {1079-7114},
  month     = sep,
  number    = {10},
  pages     = {108001},
  volume    = {101},
  doi       = {10.1103/physrevlett.101.108001},
  publisher = {American Physical Society (APS)},
}

@Article{Merifield2006,
  author    = {Merifield, R S and Sloan, S W},
  journal   = {Canadian Geotechnical Journal},
  title     = {The ultimate pullout capacity of anchors in frictional soils},
  year      = {2006},
  issn      = {1208-6010},
  month     = aug,
  number    = {8},
  pages     = {852--868},
  volume    = {43},
  doi       = {10.1139/t06-052},
  publisher = {Canadian Science Publishing},
}

@Article{Meyerhof1968,
  author    = {Meyerhof, G G and Adams, J I},
  journal   = {Canadian Geotechnical Journal},
  title     = {The Ultimate Uplift Capacity of Foundations},
  year      = {1968},
  issn      = {1208-6010},
  month     = nov,
  number    = {4},
  pages     = {225--244},
  volume    = {5},
  doi       = {10.1139/t68-024},
  publisher = {Canadian Science Publishing},
}

@Article{Rowe1982,
  author    = {Rowe, R. K. and Davis, E. H.},
  journal   = {Géotechnique},
  title     = {The behaviour of anchor plates in sand},
  year      = {1982},
  issn      = {1751-7656},
  month     = mar,
  number    = {1},
  pages     = {25--41},
  volume    = {32},
  doi       = {10.1680/geot.1982.32.1.25},
  publisher = {Emerald},
}

@Article{Shahriar2020,
  author    = {Shahriar, Azmayeen Rafat and Islam, Mohammad Shariful and Jadid, Rowshon},
  journal   = {International Journal of Geomechanics},
  title     = {Ultimate Pullout Capacity of Vertical Anchors in Frictional Soils},
  year      = {2020},
  issn      = {1943-5622},
  month     = feb,
  number    = {2},
  pages     = {-},
  volume    = {20},
  doi       = {10.1061/(asce)gm.1943-5622.0001576},
  publisher = {American Society of Civil Engineers (ASCE)},
}

@Article{Basak2023,
  author    = {Basak, Rituparna and Kozlowski, Ryan and Pugnaloni, Luis A. and Kramar, M. and Socolar, Joshua E. S. and Carlevaro, C. Manuel and Kondic, Lou},
  journal   = {Physical Review E},
  title     = {Evolution of force networks during stick-slip motion of an intruder in a granular material: Topological measures extracted from experimental data},
  year      = {2023},
  issn      = {2470-0053},
  month     = nov,
  number    = {5},
  pages     = {054903},
  volume    = {108},
  doi       = {10.1103/physreve.108.054903},
  publisher = {American Physical Society (APS)},
}

@Article{Kozlowski2021,
  author    = {Kozlowski, Ryan and Zheng, Hu and Daniels, Karen E. and Socolar, Joshua E. S.},
  journal   = {Soft Matter},
  title     = {Stress propagation in locally loaded packings of disks and pentagons},
  year      = {2021},
  issn      = {1744-6848},
  number    = {44},
  pages     = {10120--10127},
  volume    = {17},
  doi       = {10.1039/d1sm01137e},
  publisher = {Royal Society of Chemistry (RSC)},
}

@Article{Potiguar2013,
  author    = {Potiguar, Fabricio Q. and Ding, Yang},
  journal   = {Physical Review E},
  title     = {Lift and drag in intruders moving through hydrostatic granular media at high speeds},
  year      = {2013},
  issn      = {1550-2376},
  month     = July,
  number    = {1},
  pages     = {012204},
  volume    = {88},
  doi       = {10.1103/physreve.88.012204},
  publisher = {American Physical Society (APS)},
}

@Article{Wassgren2003,
  author    = {Wassgren, C. R. and Cordova, J. A. and Zenit, R. and Karion, A.},
  journal   = {Physics of Fluids},
  title     = {Dilute granular flow around an immersed cylinder},
  year      = {2003},
  issn      = {1089-7666},
  month     = Nov,
  number    = {11},
  pages     = {3318--3330},
  volume    = {15},
  doi       = {10.1063/1.1608937},
  publisher = {AIP Publishing},
}

@Article{Hilton2013,
  author    = {Hilton, J. E. and Tordesillas, A.},
  journal   = {Physical Review E},
  title     = {Drag force on a spherical intruder in a granular bed at low Froude number},
  year      = {2013},
  issn      = {1550-2376},
  month     = Dec,
  number    = {6},
  pages     = {062203},
  volume    = {88},
  doi       = {10.1103/physreve.88.062203},
  publisher = {American Physical Society (APS)},
}

@Article{Carvalho2022,
  author    = {Carvalho, Douglas D. and Lima, Nicolao C. and Franklin, Erick M.},
  journal   = {Physical Review E},
  title     = {Contacts, motion, and chain breaking in a two-dimensional granular system displaced by an intruder},
  year      = {2022},
  issn      = {2470-0053},
  month     = Mar,
  number    = {3},
  pages     = {034903},
  volume    = {105},
  doi       = {10.1103/physreve.105.034903},
  publisher = {American Physical Society (APS)},
}

@Article{Geng2005,
  author    = {Geng, Junfei and Behringer, R. P.},
  journal   = {Physical Review E},
  title     = {Slow drag in two-dimensional granular media},
  year      = {2005},
  issn      = {1550-2376},
  month     = Jan,
  number    = {1},
  pages     = {011302},
  volume    = {71},
  doi       = {10.1103/physreve.71.011302},
  publisher = {American Physical Society (APS)},
}

@Article{Wambaugh2010,
  author    = {Wambaugh, J. F. and Hartley, R. R. and Behringer, R. P.},
  journal   = {The European Physical Journal E},
  title     = {Force networks and elasticity in granular silos},
  year      = {2010},
  issn      = {1292-895X},
  month     = June,
  number    = {2},
  pages     = {135--145},
  volume    = {32},
  doi       = {10.1140/epje/i2010-10608-1},
  publisher = {Springer Science and Business Media LLC},
}

@Article{Krengel2025,
  author    = {Krengel, Dominik and Jiang, Haoran and Chen, Jian and Matsushima, Takashi},
  journal   = {Computers and Geotechnics},
  title     = {The combined effect of particle angularity and inter-particle friction on micro- and macroscopic properties of granular assemblies},
  year      = {2025},
  issn      = {0266-352X},
  month     = jan,
  pages     = {106850},
  volume    = {177},
  doi       = {10.1016/j.compgeo.2024.106850},
  publisher = {Elsevier BV},
}

@Article{Athani2019,
  author    = {Athani, Shivakumar and Rognon, Pierre},
  journal   = {Physical Review Fluids},
  title     = {Inertial drag in granular media},
  year      = {2019},
  issn      = {2469-990X},
  month     = Dec,
  number    = {12},
  pages     = {124302},
  volume    = {4},
  doi       = {10.1103/physrevfluids.4.124302},
  publisher = {American Physical Society (APS)},
}

@Article{DeGiuli2016,
  author    = {DeGiuli, E. and McElwaine, J. N. and Wyart, M.},
  journal   = {Physical Review E},
  title     = {Phase diagram for inertial granular flows},
  year      = {2016},
  issn      = {2470-0053},
  month     = July,
  number    = {1},
  pages     = {012904},
  volume    = {94},
  doi       = {10.1103/physreve.94.012904},
  publisher = {American Physical Society (APS)},
}

@Article{Li2013,
  author    = {Li, Chen and Zhang, Tingnan and Goldman, Daniel I.},
  journal   = {Science},
  title     = {A Terradynamics of Legged Locomotion on Granular Media},
  year      = {2013},
  issn      = {1095-9203},
  month     = Mar,
  number    = {6126},
  pages     = {1408--1412},
  volume    = {339},
  doi       = {10.1126/science.1229163},
  publisher = {American Association for the Advancement of Science (AAAS)},
}

@InProceedings{Li2012,
  author    = {Li, Chen and Ding, Yang and Gravish, Nick and Maladen, Ryan D. and Masse, Andrew and Umbanhowar, Paul B. and Komsuoglu, Haldun and Koditschek, Daniel E. and Goldman, Daniel I.},
  booktitle = {Earth and Space 2012},
  title     = {Toward a Terramechanics for Bio-Inspired Locomotion in Granular Environments},
  year      = {2012},
  month     = Apr,
  pages     = {264--273},
  publisher = {American Society of Civil Engineers},
  doi       = {10.1061/9780784412190.031},
}

\end{document}